\documentclass[a4]{article}

\expandafter\let\csname equation*\endcsname\undefined
\expandafter\let\csname endequation*\endcsname\undefined

\usepackage[lofdepth,lotdepth]{subfig}
\usepackage{graphicx}

\usepackage{amsmath}
\usepackage{amssymb}
\usepackage{bm}
\usepackage[a4paper,margin=22mm]{geometry}
\usepackage[T1]{fontenc}
\usepackage{lmodern}
\usepackage{hyperref}
\usepackage{microtype}
\usepackage{multicol}
\usepackage{multirow}

\newcommand{\vect}[1]{\bm{#1}}
\newcommand{\dd}{\mathrm{d}}

\usepackage{lmodern}

\usepackage{color}
\usepackage{bm}
\usepackage[normalem]{ulem}

\newcommand{\bfm}[1]{\textbf{#1}}

\newcommand{\R}[1]{{\color{black}#1}}
\newcommand{\Rs}[1]{}
\newcommand{\Rc}[1]{}
\newcommand{\Bc}[1]{}

\newcommand{\Ohmmm}{$\Omega$m$^2$}

\newcommand{\dif}{{\rm d}}
\newcommand{\dvol}{{\rm d}^3\bfm{r}}

\newcommand{\vJ}{\bfm{J}}

\newcommand{\vE}{\bfm{E}}

\newcommand{\vA}{\bfm{A}}

\newcommand{\vr}{\bfm{r}}

\newcommand{\vDJ}{\Delta\bfm{J}}
\newcommand{\vDA}{\Delta\bfm{A}}

\newcommand{\ten}[1]{\overline{\overline{#1}}}

\begin{document}

\title{A fully coupled electromagnetic-thermal-mechanical model for \R{metal-insulated} HTS high field magnets}

\author{Anang~Dadhich$^1$, Nikola~Jerance, Tara~Benkel,
				Philippe~Fazilleau$^2$ and
				Enric~Pardo$^1$$^*$\\
$^1$Institute of Electrical Engineering, Slovak Academy of Sciences,\\
Bratislava, Slovakia\\ 
$^2$Université Paris-Saclay, CEA, Département des Accélérateurs,\\
de la Cryogénie et du Magnétisme, 91191, Gif-sur-Yvette, France\\
$^*$Author to whom correspondence should be addressed (enric.pardo@savba.sk)
}

\maketitle

\begin{abstract}

Ultra high field REBCO magnets operate under strongly coupled electromagnetic, thermal and mechanical conditions, where screening currents, localized heating, thermal expansion and Lorentz forces can modify both the structural state and
the critical current density of the conductor. In this work, a coupled electromagnetic, thermal and mechanical model is developed for a metal-insulated nested REBCO insert designed for a 40 T class SuperEMFL magnet. The existing electromagnetic formulation resolves the non-uniform screening currents in the REBCO tapes. The thermal model is extended from an explicit Finite Difference Method (FDM) to an implicit Backward Euler scheme with Picard iteration, while a new axisymmetric mechanical FDM solver based on BiCGSTAB is introduced to calculate displacements, strains and stresses in the coil windings and G10 spacer regions. Thermal expansion and Lorentz force contributions are included, and the calculated longitudinal mechanical strain is coupled back to the electromagnetic model through a strain dependent \R{critical current density, which also depends on temperature, magnetic field, and its orientation.}\Rs{ $J_{\rm c}(B,T,\vartheta,\varepsilon)$ relation.} A literature-informed Parabolic-Weibull model is used to \R{model} \Rs{represent/predict} reversible and irreversible strain degradation of the REBCO conductor. The numerical methods are benchmarked, and the resulting framework provides a computationally efficient approach for investigating temperature gradients, thermo-mechanical stresses, strain-dependent critical current degradation and quench behaviour in full scale nested high field REBCO magnets.

\end{abstract}

\section{Introduction}

REBCO high temperature superconducting (HTS) coated conductors are one of the main enabling technologies for superconducting magnets above 30 T because they retain a high engineering current density at magnetic fields where conventional Low Temperature Superconductors (LTS) become strongly limited. Several REBCO insert magnets have demonstrated this capability at full or near full scale. The Little Big Coil (LBC) generated 14.4 T inside a 31.1 T resistive background field, producing a record 45.5 T total DC field \cite{hahnS2019Nat}. In Europe, the metal insulated Nougat REBCO insert reached a total field of 32.5 T at 4.2 K inside a resistive background magnet \cite{lecrevisseT2022SSTa}. More recently, the MIT 1.3 GHz NMR program constructed and tested the H835 single coil REBCO insert, designed to
generate 19.6 T together with an approximately 11 T LTS outsert towards a
total field of 30.5 T \cite{dongF2025SST}. Other high field facilities and
hybrid magnet systems have also demonstrated or targeted magnetic fields
above 30 T \cite{pugnatP2022IES}.

The next generation of high field magnets aims to replace the
power intensive resistive background magnet by an LTS outsert while retaining
a compact REBCO insert. In the SuperEMFL project, fully superconducting
systems generating 32 T and more than 40 T are being developed using metal
insulated REBCO windings inside LTS background magnets \cite{superEMFL, dadhichA2026SUSTa, durochatM2024IES, fazilleauP2024IES}.
For the 40 T class system, both single solenoid and nested REBCO insert
configurations have been investigated \cite{durochatM2024IES}. The nested
design provides additional freedom in distributing magnetic field, current
density and stored energy between several coaxial HTS coils. However, these
coils are strongly coupled, and a local disturbance can redistribute current,
heat and electromagnetic forces through the complete magnet
\cite{fazilleauP2024IES, dadhichA2024SSTa, pardoE2026RIE}.

Operation at such high magnetic fields introduces closely coupled thermal and
mechanical limitations. Large transport currents and magnetic fields generate
substantial Lorentz forces and consequently hoop, radial, axial and shear
stresses within the winding pack. Screening currents in the wide REBCO tapes
further modify the local current and force distributions
\cite{yanY2020SST, zhouB2023PhyC, wangYN2024SST, suB2025JAP}. At the same
time, AC losses during magnet charging, local defects, current sharing and
operation close to the critical current can generate localized heating.
Because normal zone propagation in REBCO conductors is relatively slow, this
heat can remain concentrated and develop into large temperature gradients or
thermal runaway \cite{badelA2019SST, gavrilinAV2024IES, pardoE2026RIE}.
Nonuniform thermal expansion then introduces additional mechanical stresses,
while excessive strain can reduce the critical current density and eventually
lead to irreversible conductor damage. The thermal, electromagnetic and
mechanical responses of ultra high field REBCO magnets are therefore strongly
interdependent.

Existing numerical models generally describe only part of this coupled
behaviour. Detailed electromagnetic models can resolve screening currents,
while mechanical models are most commonly based on the Finite Element Method
(FEM) and often receive the electromagnetic force distribution from a
separate calculation \cite{leeJ2018IEEE, berrospeE2020IEEE, yanY2020SST, shaoL2021Ele, zhouB2023PhyC, suB2025JAP}. Conversely, many electrothermal quench models use uniform or reduced current distributions to obtain practical computation
times \cite{wangY2022PhyC, badelA2019SST, dongF2022APL, vitranoA2023IES},
which can neglect screening current losses and their nonuniform force
distribution \cite{dadhichA2024SSTa, pardoE2026RIE}. More complete
electromagnetic, thermal and mechanical models have recently been reported
for REBCO coils \cite{wangYN2024SST}, while equivalent circuit and PEEC
approaches provide efficient descriptions of current redistribution in
noninsulated and metal insulated windings
\cite{noguchiS2022SST, changZ2025ATE}. Nevertheless, simultaneously resolving
screening currents, transient temperature gradients, thermo mechanical
stresses and strain dependent $J_{\rm c}$ in a full scale nested insert
remains computationally demanding\R{, especially for metal-insulated or non-insulated windings}.

For the thermal model, the present formulation extends the explicit Finite
Difference Method (FDM) previously developed by the authors for REBCO coils
and full scale high field magnets \cite{pardoE2023IES, dadhichA2024SSTa, hussainA2024SST}. The explicit formulation was computationally inexpensive but required small thermal time steps imposed by the diffusion stability condition. Here, this limitation is removed by introducing an implicit Backward Euler discretization with Picard iteration for temperature dependent material properties and heat sources.
FDM thermal models have also been applied to low resistance and noninsulated REBCO pancake coils \cite{Markiewicz2016, Markiewicz2019}, and Genot \emph{et al.} coupled a PEEC electrical model to a two dimensional FDM thermal solver for multipancake coils \cite{Genot2022}. More recently, an
implicit ADI thermal solver using a finite volume spatial discretization was
applied to an NI REBCO magnet \cite{Zheng2024}. For the regular axisymmetric
$(r,z)$ winding geometry considered here, FDM provides a compact conservative
stencil with low storage and avoids complex mesh generation and matrix
assembly. The present thermal formulation therefore extends the authors'
previous explicit approach to an implicit solver suitable for the coupled
electromagnetic, thermal and mechanical model.

For the mechanical problem, a cell centred FDM solver using the
Bi Conjugate Gradient STABilized (BiCGSTAB) iterative method is developed.
Mechanical analysis of REBCO magnets is usually performed using FEM, whereas
finite difference and BiCGSTAB methods have been used mainly in other
superconducting or numerical applications
\cite{Du2005, CalhounLopez2001, CostaBouzo2004}. The present formulation
provides compact stencils, matrix free operator evaluation and low storage on
the same axisymmetric mesh already used by the thermal model. BiCGSTAB is
suitable for the resulting large and generally nonsymmetric linear system and
avoids the symmetry requirement of conjugate gradient methods
\cite{vanDerVorst1992, Saad2003}. Although FDM provides less geometrical
flexibility than FEM, its use of the existing mesh and numerical framework
allows an efficient mechanical solution to be incorporated directly into the
multiphysics model.

The present work applies this framework to the metal insulated nested REBCO
insert developed for the SuperEMFL 40 T class magnet
\cite{durochatM2024IES, superEMFL}. The configuration contains two concentric
HTS stacks operating inside an approximately 15 T LTS background field.
The electromagnetic model resolves the nonuniform current density and
screening currents in the REBCO tapes, while the new implicit thermal and
mechanical FDM solvers calculate the transient temperature, displacement,
strain and stress distributions on the same axisymmetric mesh. The mechanical
response includes thermal expansion and Lorentz force contributions, and the
calculated longitudinal mechanical strain is coupled back to the
electromagnetic model through the strain dependent
$J_{\rm c}(B,T,\vartheta,\varepsilon)$ relation. The aim is therefore to
provide a computationally efficient coupled electromagnetic, thermal and
mechanical framework for studying the operating limits and quench behaviour
of full scale nested high field REBCO magnets.

\begin{figure}[tbp]
	\centering

{\includegraphics[trim=0 0 0 0,clip,width=12 cm]{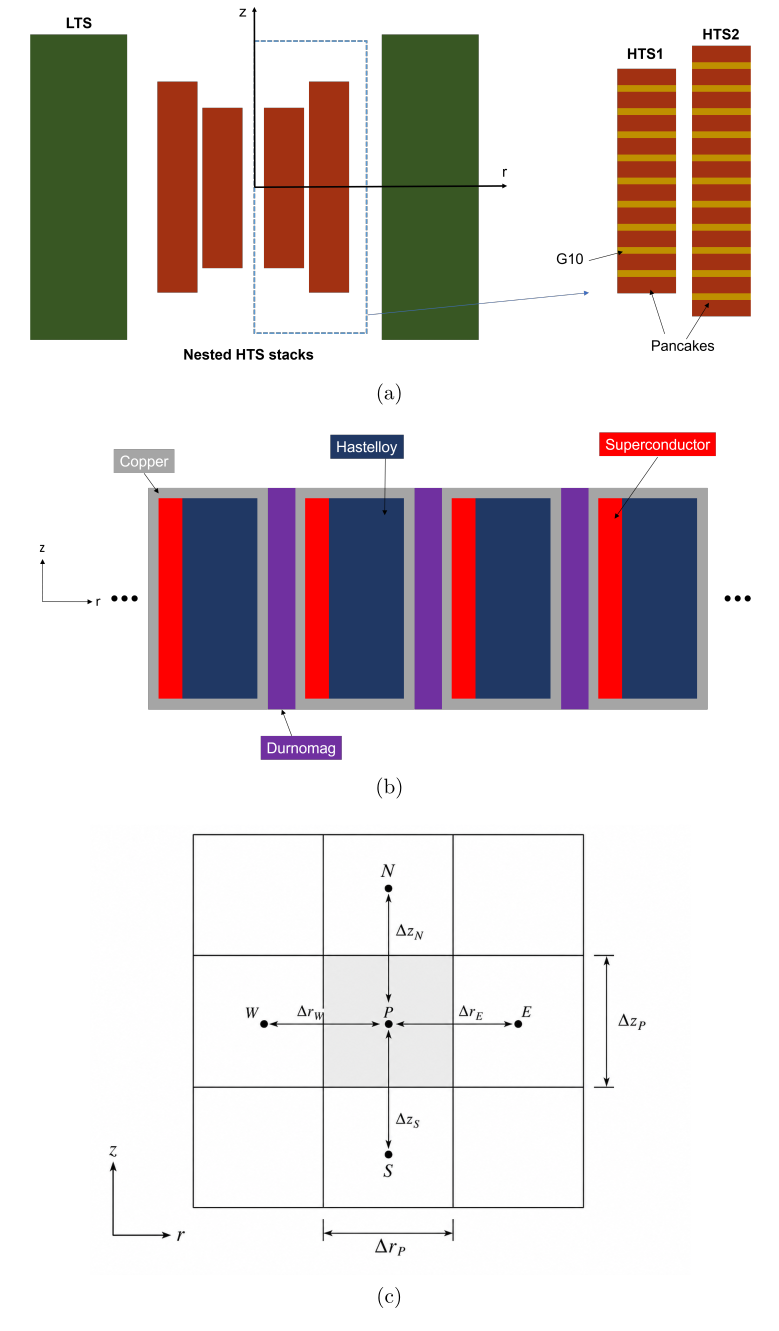}}%

\caption{(a) Full magnet and its cross section, which shows nested HTS1 and HTS2 pancake coil insert stacks inside LTS outsert magnet. (b) Cross section of turns in pancakes, including the isolating Durnomag layer between 2 tapes. (c) shows localized mesh grid for Finite Difference discretization for cell P and the neighboring cells. The sketches are only for representation and not to scale.  }
\label{crossSec}
\end{figure}

\section{Complete multiphysics model}

We have developed a complete multiphysics model, which fully couples electromagnetic, electrothermal, and mechanical behaviour of a group of pancake coils in a high field magnet, for example, the studied case of nested stacks in Fig. \ref{crossSec}. These separate modules of electromagnetic, thermal, and mechanical responses for 2D axisymmetric coils, as well as the model coupling and material homogenization are explained in subsections below.

\subsection{Strain dependence of the critical current density}
\label{s.JcBTe}

\begin{figure*}[tbp]
  \centering

{\includegraphics[trim=0 0 0 0,clip,width=14 cm]{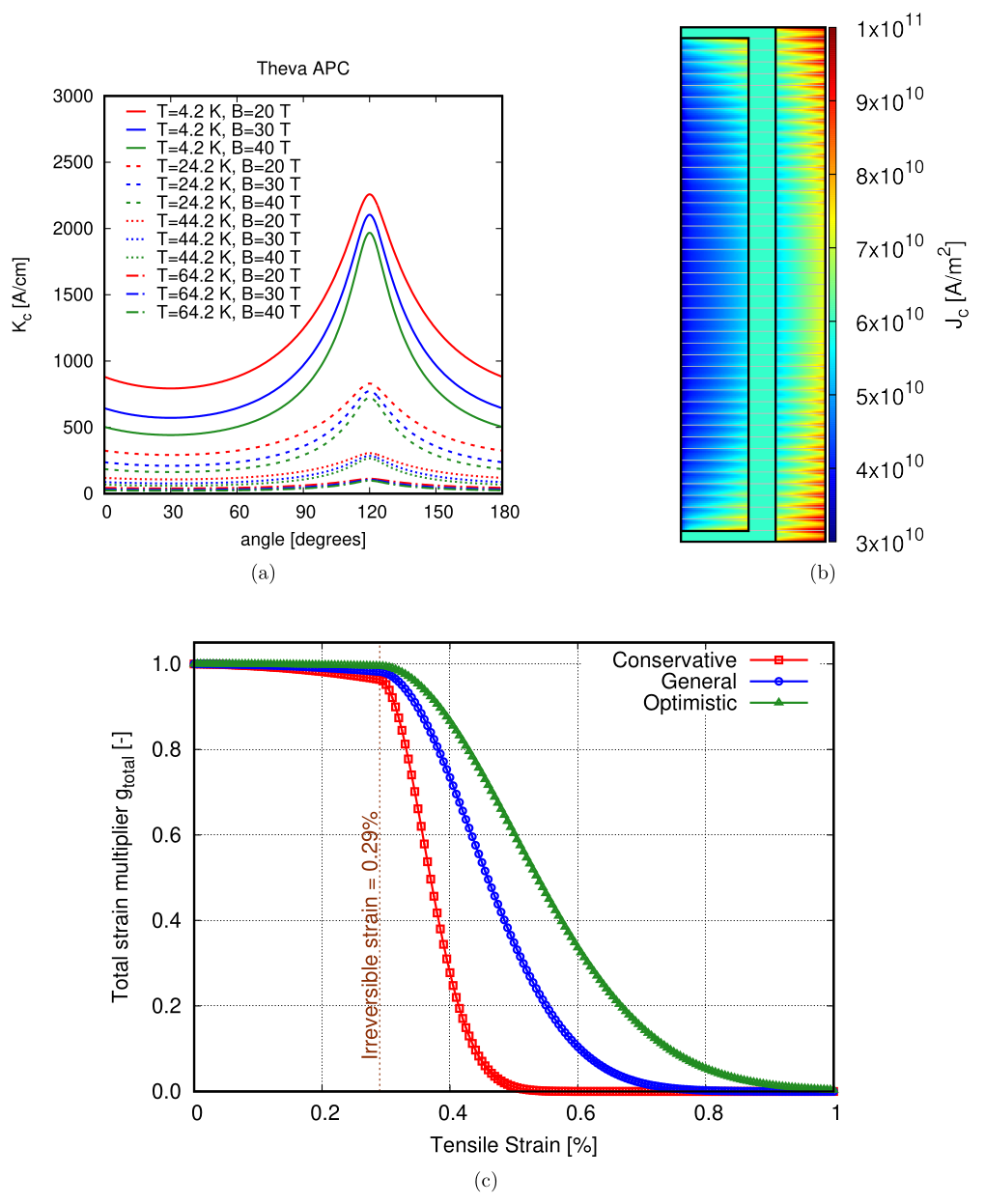}}%

  \caption{The superconducting Theva APC presents asymmetrical angular dependence of the tape critical current per unit tape width, $K_c$ ($J_c=K_c/d$, where $d$ is the superconducting layer thickness). (a) The shown dependence on angle, temperature and magnetic field are from the fits in \cite{pardoE2026RIE}, which are based on experimental data. (b) shows the considered symmetric $J_c$ achieved after ramp up. (c) shows the $g_{total}$ multiplier as defined in mechanical model according to which the $J_c(\varepsilon)$ strain dependence behaves. }
  \label{JcBT}
\end{figure*}

The mechanical feedback to the electromagnetic model is introduced through a dimensionless strain multiplier applied to the reference critical current density ($J_{\rm c}$) law. The strain obtained directly from the displacement field represents the total geometric strain. In particular, the total geometric hoop strain is
\begin{equation}
\varepsilon_{\theta\theta}^{\rm tot}
=
\frac{u_r}{r} = \varepsilon_{\theta\theta}^{\rm mech} + \varepsilon_{\theta\theta}^{\rm th},
\end{equation}
where, $\varepsilon_{\theta\theta}^{\rm mech}$ and $\varepsilon_{\theta\theta}^{\rm th}$ are longitudinal strains due to mechanical configuration (given constraints) and free thermal expansion, respectively. 
A local temperature change produces a free thermal strain,
\begin{equation}
\varepsilon_{\theta\theta}^{\rm th}
=
\alpha_\theta \left(T-T_i\right),
\end{equation}
where $T_i$ is the temperature corresponding to the mechanical reference configuration, and $\alpha_\theta$ is hoop thermal expansion coefficient. The mechanical hoop strain acting on the conductor is
therefore
\begin{equation}
\label{e.ett_jc}
\varepsilon_{\theta\theta}^{\rm mech}
=
\varepsilon_{\theta\theta}^{\rm tot}
-
\varepsilon_{\theta\theta}^{\rm th}
=
\frac{u_r}{r}
-
\alpha_\theta\left(T-T_i\right).
\end{equation}
Thus, a freely expanding conductor has
$\varepsilon_{\theta\theta}^{\rm mech}=0$, whereas any restriction of its free thermal expansion produces a non-zero mechanical strain. The quantity $\varepsilon_{\theta\theta}^{\rm mech}$ is therefore used as the longitudinal strain
input to the strain-dependent $J_c$ model, while the direct temperature dependence is independently included through
$J_c^0(B,T,\vartheta)$, where $J_{\rm c}^{0}$ is the strain-reference electromagnetic law that is dependent on magnetic field ($B$), temperature ($T$), and field angle ($\vartheta$), as shown in Fig.~\ref{JcBT}(a).

The longitudinal strain of the REBCO layer is approximated by the hoop strain obtained from the axisymmetric mechanical
solution, $\varepsilon\simeq\varepsilon_{\theta\theta}^{\rm mech}$ (shortened terminology we use in this section for relevant strain). The degradation of $J_{\rm c}$ under longitudinal hoop strain is modeled using a compound
Parabolic-Weibull (PW) framework. The parabolic component describes the intrinsic reversible variation of the critical current with longitudinal strain, while the Weibull component represents the irreversible loss of current-carrying capability after mechanical damage develops in the superconducting layer \cite{vanDerLaan2011,Gao2020,Arai2010,Ochiai2012}. The coupled $J_{\rm c}$ law is written as

\begin{equation} \label{e.jc_strain_master_pw}
 { J_{\rm c}(B,T,\vartheta,\varepsilon, \varepsilon_{\rm max} )
 =J_{\rm c}^{0}(B,T,\vartheta)\,g_{\rm rev}(\varepsilon)\,g_{\rm W}(\varepsilon_{\rm max}) },
\end{equation}

Here, $g_{\rm rev}$ and
$g_{\rm W}$ are the reversible and Weibull irreversible multipliers as described below, respectively, and $\varepsilon_{\rm max}$ denotes the maximum achieved strain.

The present coupling represents \emph{electrical} degradation only.  Thus, no plastic deformation is assumed for the homogenized coil cross-section, although Jc degradation inherently assumes plastic deformation and cracking of the superconducting layer.The factor $g_{\rm W}$ reduces the electrically effective current-carrying
capability of the REBCO layer, but it does not modify the elastic constants,
yield surface, plastic strain, geometry, contact state or crack opening of the
winding pack. Consequently, $D=1-g_{\rm W}$ may be interpreted as a
homogenized loss of electrically effective superconducting area. This interpretation is consistent with
coated-conductor experiments in which irreversible $I_{\rm c}$ degradation is
associated with cracking of the superconducting layer and redistribution of
the transport current around damaged regions
\cite{Arai2010,Cheggour2005,Ochiai2012}.

For the considered THEVA Pro-Line conductor, a complete experimental
$J_{\rm c}(\varepsilon,B,T,\vartheta)$ characterization covering the
high-field and temperature range of interest is not presently available.
The strain dependence is therefore constructed from THEVA-specific
measurements where available and from electromechanical data and
phenomenological models reported for comparable commercial REBCO conductors
\cite{Arai2010,SuperEMFLTARSIS,Barth2015,Zhou2016,Pierro2019,Gao2020}.
A numerical sensitivity analyis is thereby performed to assess the expected Theva $J_c(\varepsilon)$ behavior, and thus three literature-informed parameter sets are retained:
Conservative, General and Optimistic. They represent progressively stronger,
nominal and weaker strain degradation, respectively, and are intended as
engineering bounds rather than statistically calibrated confidence limits
for the investigated tape. The corresponding parameters are listed in
Table~\ref{t.theva_strain_parameter_sets}.

The $J_c(\varepsilon)$ dependence thus achieved from these parameters is shown in Figure \ref{JcBT} (c), which is the result of the multiplication of $g$ multipliers to the maximum $Jc$ in Figure \ref{JcBT} (a). The Conservative and General curves that we achieve using sensitivity analysis agrees closely with $J_c(\varepsilon)$ dependence measured for 2 different Theva samples in SuperEMFL project \cite{SuperEMFLTARSIS}, and thus validates our process.

\begin{table}[ht]
\centering
\caption{THEVA-oriented parameter sets used for the strain-dependent
critical-current analysis. All strain quantities are expressed as
dimensionless fractions.}
\label{t.theva_strain_parameter_sets}

\small
\renewcommand{\arraystretch}{1.15}

\begin{tabular}{lccc}
\hline
Parameter
& Conservative
& General
& Optimistic \\
\hline

$a$
& $9378$
& $4667$
& $1300$ \\

$\varepsilon_m$
& $-0.0004$
& $-0.0004$
& $-0.0004$ \\

$n_{\varepsilon}$
& $2.18$
& $2.18$
& $2.18$ \\

$\varepsilon_{\mathrm{ref}}$
& $0$
& $0$
& $0$ \\

$\varepsilon_{\mathrm{irr}}$
& $0.0029$
& $0.0029$
& $0.0029$ \\

$\lambda_{\varepsilon}$
& $0.0010$
& $0.0021$
& $0.0030$ \\

$m$
& $2.0$
& $2.0$
& $2.0$ \\

\hline
\end{tabular}
\end{table}

\subsubsection{The Parabolic-Weibull (PW) Model}

In the PW model, the total strain multiplier, $g_{\rm total}$ (Figure \ref{JcBT} (c)), is defined as
the product of a reversible parabolic function, $g_{\rm rev}$, and an
irreversible Weibull survival function, $g_{\rm W}$:

\begin{equation} \label{e.gmultiplier}
g_{\rm total}(\varepsilon,\varepsilon_{\max})
=
g_{\rm rev}(\varepsilon)\,
g_{\rm W}(\varepsilon_{\max}).
\end{equation}

The parameter $\varepsilon_{\rm ref}$ denotes the strain state represented by
the reference electromagnetic law $J_{\rm c}^{0}(B,T,\vartheta)$.
Fabrication or thermal pre-strains that are already represented by the experimental $J_{\rm c}^{0}$ data are
therefore not added again, avoiding double counting of the strain state. The reversible component governs the temporary variation of the current-carrying capability with elastic longitudinal strain and is formulated
as \cite{vanDerLaan2011}

\begin{equation} \label{e.gmultiplierrev}
g_{\rm rev}(\varepsilon)
=
\frac{
1-a|\varepsilon-\varepsilon_m|^{n_{\varepsilon}}
}{
1-a|\varepsilon_{\rm ref}-\varepsilon_m|^{n_{\varepsilon}}
}.
\end{equation}

Here, $a$ is the reversible strain-curvature coefficient,
$\varepsilon_m$ is the strain state at which the reversible critical current
is maximum, $n_{\varepsilon}$ is a power-law exponent, and
$\varepsilon_{\rm ref}$ normalizes the curve to the reference electromagnetic
data. The exponent $n_{\varepsilon}=2.18$ is retained from the full-tape REBCO
fits reported by van der Laan \emph{et al.}
\cite{vanDerLaan2011}.

Irreversible permanent degradation, activated when the historical maximum
tensile strain $\varepsilon_{\max}$ exceeds the damage-onset threshold
$\varepsilon_{\rm irr}$, is represented using a Weibull survival function
\cite{Gao2020},

\begin{equation} \label{e.gmultiplierW}
g_{\rm W}(\varepsilon_{\max})
=
\exp
\left[
-
\left(
\frac{
\max(0,\varepsilon_{\max}-\varepsilon_{\rm irr})
}{
\lambda_{\varepsilon}
}
\right)^m
\right],
\end{equation}

where $\lambda_{\varepsilon}$ is the excess-strain scale controlling the
post-onset degradation rate and $m$ is the Weibull shape parameter.

The considered THEVA Pro-Line conductors use an
inclined-substrate-deposition (ISD) architecture, with an oriented MgO buffer
on a Hastelloy C-276 substrate and a GdBaCuO superconducting layer
\cite{Prusseit2005,THEVAProLine}. Earlier electromechanical studies of THEVA
conductors investigated closely related DyBCO/MgO/Hastelloy ISD architectures
\cite{Arai2010,Ochiai2012}. These measurements provide useful
manufacturer- and architecture-specific evidence, but they do not constitute
a complete $J_{\rm c}(B,T,\vartheta, \varepsilon)$ fit for the present
Pro-Line tape.

\subsubsection{Anchoring the Irreversible Damage Threshold
($\varepsilon_{\rm irr}=0.0029$)}

In this work, the irreversible damage onset threshold is fixed at
$\varepsilon_{\rm irr}=0.29\%$ ($0.0029$ strain) across all three global
scenarios (Conservative, General, and Optimistic). This value is selected
directly from the SuperEMFL characterization of commercial THEVA sample
240019, for which an irreversible strain limit of approximately $0.29\%$
was reported under tensile loading \cite{SuperEMFLTARSIS}. A second THEVA
sample reported in the same study showed a comparable irreversible strain
limit, supporting the use of this value as a THEVA-specific experimental
anchor.

By fixing $\varepsilon_{\rm irr}$ across the three scenarios, the experimentally
anchored onset of permanent $I_{\rm c}$ degradation is kept constant while
the uncertainty in the subsequent damage evolution is varied independently.
The parameter $\varepsilon_{\rm irr}$ is therefore interpreted as the onset
of irreversible electrical degradation and not as the mechanical yield strain
of the superconducting layer.

\subsubsection{Role of the Post-Onset Scale Parameter
($\lambda_{\varepsilon}$)}

While the onset of irreversible degradation is fixed by
$\varepsilon_{\rm irr}$, the rate at which the current-carrying capability
decreases after this threshold depends on the subsequent development of
mechanical damage. This behavior can depend strongly on conductor
architecture, stabilizer thickness, interfaces and reinforcement, which are
known to influence the irreversible strain tolerance of coated conductors
\cite{Cheggour2005,Ochiai2012,Barth2015,Zhou2016}.

The parameter $\lambda_{\varepsilon}$ controls the characteristic excess
strain over which the Weibull multiplier decreases after
$\varepsilon_{\rm irr}$ is exceeded. The central value
$\lambda_{\varepsilon}=0.21\%$ is adopted from the experimentally fitted
REBCO electromechanical model of Gao \emph{et al.} \cite{Gao2020}.
Because an equivalent post-onset scale has not been reported for the
considered THEVA tape, values of $0.10\%$ and $0.30\%$ are used as lower
and upper sensitivity bounds.

The three global cases therefore represent distinct post-onset degradation
regimes:
\begin{enumerate}
    \item \textbf{Conservative ($\lambda_{\varepsilon}=0.10\%$):}
    represents rapid post-onset degradation. The relatively small
    characteristic excess strain produces a fast reduction of $g_{\rm W}$
    after the irreversible threshold and is used as the engineering lower
    bound.

    \item \textbf{General ($\lambda_{\varepsilon}=0.21\%$):}
    uses the characteristic strain scale obtained from the commercial-REBCO
    electromechanical fit of Gao \emph{et al.} \cite{Gao2020} and is adopted
    as the central analogue for the present THEVA analysis.

    \item \textbf{Optimistic ($\lambda_{\varepsilon}=0.30\%$):}
    represents slower post-onset degradation and therefore greater retained
    current-carrying capability at a given excess strain. This upper bound is
    consistent with the general observation that stabilization and conductor
    architecture can increase strain tolerance, although
    $\lambda_{\varepsilon}=0.30\%$ is not treated here as a fitted THEVA
    material parameter \cite{Cheggour2005,Ochiai2012}.
\end{enumerate}

\subsubsection{Bounding the Reversible Response
($a$}

The reversible variation of $J_{\rm c}$ originates from
the strain sensitivity of the superconducting state and associated changes
in the REBCO electromechanical response
\cite{vanDerLaan2011,Cheggour2005,Pierro2019}.
To account for the uncertainty associated with the reversible behavior of the
considered THEVA conductor, the curvature of the PW relation is varied across
the three parameter sets.

The parabolic curvature parameter $a$ is varied from $1300$ (Optimistic) to
$4667$ (General) and $9378$ (Conservative). These values span the magnitude
of published power-law fits for commercial REBCO conductors
\cite{vanDerLaan2011,Gao2020}. The General value,
$a=4667$, together with $n_{\varepsilon}=2.18$ and
$\varepsilon_m=-0.04\%$, corresponds to the full-tape GdBCO S-1 analogue
reported by van der Laan \emph{et al.} \cite{vanDerLaan2011}.
For the Conservative and Optimistic cases, $n_{\varepsilon}$ and
$\varepsilon_m$ are deliberately held at their General values while only
$a$ is changed. These two combinations therefore represent
literature-informed curvature bounds rather than complete published conductor
fits.

\subsubsection{Controlled Parameters
($m$, $n_{\varepsilon}$, $\varepsilon_m$, $\varepsilon_{\rm ref}$)}

To isolate the effects of the principal sensitivity parameters and maintain a
clear ordering between the Conservative, General and Optimistic engineering
scenarios, several coefficients are held fixed.

The Weibull modulus is fixed at $m=2.0$, consistent with the
commercial-REBCO electromechanical fit of Gao \emph{et al.}
\cite{Gao2020}. Varying $m$ changes the shape of the post-onset degradation
curve and can alter the relative ordering of the different scenarios at
different excess strains; fixing it therefore allows the effect of
$\lambda_{\varepsilon}$ to be assessed independently.

The power-law exponent $n_{\varepsilon}=2.18$ and peak-strain position
$\varepsilon_m=-0.04\%$ ($-0.0004$) are retained from the GdBCO S-1
analogue reported by van der Laan \emph{et al.}
\cite{vanDerLaan2011}, maintaining a consistent reversible reference across
all simulations. Finally, $\varepsilon_{\rm ref}=0$ defines the strain datum
associated with the adopted reference $J_{\rm c}^{0}$ law; it does not imply
that the superconducting layer is microscopically free of residual strain.

For final accurate PW model parameters for Theva tapes, very detailed mechanical measurements are required that range from various temperatures to background magnetic fields, that can give these actual parameters from experimental fits. However, as the shown curves for General and Conservative cases closely resemble the $J_c(\varepsilon)$ behavior from SuperEMFL deliverables \cite{SuperEMFLTARSIS}, we consider them as good approximation for this initial study and model completion.

\subsection{Electromagnetic model}
\label{s.EM}

In this article, we use the Minimum Electro-Magnetic Entropy Production (MEMEP) method \cite{pardoE2017JCP} to compute the electromagnetic quantities for metal-insulated and non-insulated magnets, as detailed in \cite{pardoE2024SSTa, pardoE2026RIE}. 

Summarizing, MEMEP solves the following integral equation for $\vJ$
\begin{equation} \label{e.EAphi}
\vE(\vJ)=-\frac{\partial \vA[\vJ]}{\partial t}-\frac{\partial \vA_a}{\partial t}-\nabla\phi,
\end{equation}
with
\begin{equation}
\vA[\vJ](\vr)=\frac{\mu_0}{4\pi}\int_\Omega\dvol' \frac{\vJ(\vr')}{|\vr-\vr'|}.
\end{equation}
Above, we used Coulomb's gauge for $\vA$, and hence $\nabla\cdot\vA=0$ and $|\vA|\to 0$ at $|\vr|\to\infty$ \cite{pardoE2023book}. In (\ref{e.EAphi}), $\vA_a$ is the applied vector potential (a vector potential due to external sources), and $\phi$ is the electrostatic potential. As demonstrated in \cite{pardoE2017JCP}, solving this equation for a given time $t$ can be done by minimizing the following functional, provided that we know the solution at time $t-\Delta t$, $J(t-\Delta t)$, being $\Delta t$ a certain time interval
\begin{eqnarray} \label{L3D}
L[\vJ] & = & \int_\Omega\dvol \biggl\{ \frac{1}{2}\vDJ\cdot\frac{\vDA[\vDJ]}{\Delta t} + \vDJ\cdot\vDA_a \nonumber \\
&& + U(\vJ) +\nabla\phi\cdot\vJ \biggr\},
\end{eqnarray}
where $U(\vJ)$ is defined as
\begin{equation}
U(\vJ)\equiv \int_0^{\vJ}\dif\vJ'\cdot \vE(\vJ').
\end{equation}
Above, $\Delta\vJ = \vJ(t)-\vJ(t-\Delta t)$, $\Delta\vA_a = \vA_a(t)-\vA_a(t-\Delta t)$, and $\vE(\vJ)$ is any nonlinear relation, such as that in \cite{pardoE2026RIE}.

For non-insulated and metal-insulated coils, axial symmetry can be implemented by using the additional constraint between the total input current $I$ and the angular and radial current in each turn $i$, $I_{\varphi i}$ and $I_{ri}$, as detailed in \cite{pardoE2024SSTa}
\begin{equation}
I=I_{\varphi i}+I_{ri}.
\end{equation}

\subsection{Thermal model: Implicit FDM}
\label{sec:thermal_fd}

The transient temperature field is solved in the axisymmetric $(r,z)$ domain using a fully implicit finite-difference scheme, while in our previous works we used explicit finite-differences \cite{dadhichA2024SSTa, dadhichA2026SUSTa}. Under axisymmetry, $\partial/\partial\theta=0$, and the effective thermal conductivity tensor is taken as diagonal in the principal material directions, $\ten{k}={\rm diag}(k_r,k_\theta,k_z)$. The governing heat equation therefore reduces to \cite{CarslawJaeger1959}

\begin{equation} \label{e.thermal_axisym}
 { c_p(T)\frac{\partial T}{\partial t}
 = \frac{1}{r}\frac{\partial}{\partial r}
 \left[r k_r(T)\frac{\partial T}{\partial r}\right]
 +\frac{\partial}{\partial z}
 \left[k_z(T)\frac{\partial T}{\partial z}\right]
 +p },
\end{equation}

where $T=T(r,z,t)$ is the temperature, $c_p(T)$ is the effective volumetric heat capacity, $k_r(T)$ and $k_z(T)$ are the effective radial and axial thermal conductivities, respectively, and $p$ is the volumetric heat-generation density. In the coupled model, $p$ contains the electromagnetic dissipation supplied by the MEMEP solution.

\subsubsection{Spatial grid and conservative finite differences}

A cell-centred grid is used with spacings $\Delta r$ and $\Delta z$, which can be non-uniform in mesh. The temperature at the centre of cell $(i,j)$ is denoted by $T_{i,j}$, and its radial and axial coordinates are, respectively,

\begin{equation} \label{e.thermal_grid_r}
 { r_i=r_{\min}
 +\sum_{\ell=1}^{i-1}\Delta r_\ell
 +\frac{\Delta r_i}{2} }.
\end{equation}

\begin{equation} \label{e.thermal_grid_z}
 { z_j=z_{\min}
 +\sum_{\ell=1}^{j-1}\Delta z_\ell
 +\frac{\Delta z_j}{2} }.
\end{equation}

For the derivation below, the cell $(i,j)$ is denoted by $P$, while its neighbours $(i+1,j)$, $(i-1,j)$, $(i,j+1)$ and $(i,j-1)$ are denoted by $E$, $W$, $N$ and $S$, respectively, as shown in Figure \ref{crossSec} (c). The radial east and west , and axial north and south face locations are

\begin{equation} \label{e.radial_faces}
 { r_e=r_i+\frac{\Delta r_P}{2},
 \qquad
 r_w=r_i-\frac{\Delta r_P}{2} ,
 \qquad
 z_n=z_j+\frac{\Delta z_P}{2} ,
 \qquad
 z_s=z_j-\frac{\Delta z_P}{2} },
\end{equation}
where, $\Delta r_E$, $\Delta r_W$, $\Delta z_N$, and $\Delta z_S$ are distances of point P from neighboring east, west, north, and south mesh elements as shown in Figure \ref{crossSec} (c), such that

\begin{eqnarray}
&& \Delta r_E=r_{i+1}-r_i,
\qquad
\Delta r_W=r_i-r_{i-1}, \\
&& \Delta z_N=z_{j+1}-z_j,
\qquad
\Delta z_S=z_j-z_{j-1}.
\label{e.thermal_neighbour_distances}
\end{eqnarray}

The diffusion terms are discretized in conservative flux form. This preserves continuity of heat flux across neighbouring cells and is particularly useful when the conductivity changes strongly between winding, insulation, spacer and structural regions. On an orthogonal cell-centred grid, this conservative flux discretization gives the same five-point algebraic structure commonly used in finite-difference heat solvers \cite{Patankar1980,Moukalled2016, dadhichA2026SUSTa}. The radial contribution at $P$ is written as

\begin{equation} \label{e.radial_fd_flux}
 { \left.\frac{1}{r}\frac{\partial}{\partial r}
 \left(rk_r\frac{\partial T}{\partial r}\right)\right|_P
 \simeq
 \frac{1}{r_i\Delta r_P}
 \left[
 r_e k_{r,e}\frac{T_E-T_P}{\Delta r_E}
 -r_w k_{r,w}\frac{T_P-T_W}{\Delta r_W}
 \right] }.
\end{equation}

Similarly, the axial contribution is

\begin{equation} \label{e.axial_fd_flux}
 { \left.\frac{\partial}{\partial z}
 \left(k_z\frac{\partial T}{\partial z}\right)\right|_P
 \simeq
 \frac{1}{\Delta z_P}
 \left[
 k_{z,n}\frac{T_N-T_P}{\Delta z_N}
 -k_{z,s}\frac{T_P-T_S}{\Delta z_S}
 \right] }.
\end{equation}

Here, $\Delta r_P$ and $\Delta z_P$ are current cell P's radial and axial thicknesses, respectively. Also, $k_{r,e}$ and $k_{r,w}$ are the effective radial conductivities at the east and west faces, whereas $k_{z,n}$ and $k_{z,s}$ are the corresponding axial face conductivities. For smooth material properties, Eqs.~(\ref{e.radial_fd_flux}) and (\ref{e.axial_fd_flux}) are second-order accurate in space on a uniform grid. The effective conductivities can also be calculated at the edge of coils where conduction or convection cooling occurs, as derived in our previous work \cite{dadhichA2024SSTa}.

\subsubsection{Backward-Euler time integration and Picard linearization}

Let $t^n$ and $t^{n+1}=t^n+\Delta t$ denote two consecutive time steps, where $\Delta t$ is the time step size, which does not need to be constant. Then, the temperature time derivative is approximated using the first-order Backward-Euler scheme,

\begin{equation} \label{e.backward_euler}
 { \left.\frac{\partial T}{\partial t}\right|_P^{n+1}
 \simeq
 \frac{T_P^{n+1}-T_P^n}{\Delta t} }.
\end{equation}

Backward Euler is first-order accurate in time and unconditionally stable for the linear diffusion equation, which makes it attractive for strongly diffusive cryogenic thermal problems and for coupling to electromagnetic time steps \cite{LangtangenLinge2017,MortonMayers2005}. The resulting problem is nevertheless nonlinear when $c_p$, $k_r$, or $k_z$ depend on temperature. Implicit-Euler finite-difference formulations are commonly used for such nonlinear heat-conduction problems \cite{FilipovFarago2018}.

The nonlinear coefficients are treated by Picard fixed-point iteration. At
time level $n+1$, the superscript $(m)$ denotes the current Picard iterate and
$(m+1)$ the updated iterate. The Picard iteration starts from an initial
temperature estimate, taken here as the converged temperature from the
previous time step,

\begin{equation}
 { T_P^{n+1,(0)}=T_P^n }.
\end{equation}

Thus, at the first Picard iteration ($m=0$), the heat capacity and thermal
conductivities are evaluated using $T_P^n$. These coefficients are then kept
fixed while solving the linear system for the first updated estimate
$T_P^{n+1,(1)}$. The material properties are subsequently recomputed using
this new temperature field and the procedure is repeated for
$m=1,2,\ldots$ until convergence. The temperature-dependent material
properties are therefore lagged according to

\begin{eqnarray}
&& c_{p,P}^{(m)}=c_p\!\left(T_P^{n+1,(m)}\right), \\
&& k_{r,P}^{(m)}=k_r\!\left(T_P^{n+1,(m)}\right),
\qquad
k_{z,P}^{(m)}=k_z\!\left(T_P^{n+1,(m)}\right).
\label{e.picard_coefficients}
\end{eqnarray}

The electromagnetic heat-generation density $p_P$ is obtained from the
coupled electromagnetic solution for the current time step and is held
constant throughout the Picard iterations. The face conductivities are then
recomputed from the lagged cell-centred conductivities using the formulas
developed in \cite{dadhichA2024SSTa}.

With these coefficients fixed, the equation for $T^{n+1,(m+1)}$ is linear. After solving this system, the newly obtained temperature $T^{n+1,(m+1)}$ becomes the temperature estimate used to evaluate the coefficients in the next Picard iteration. This fixed-point treatment is the standard Picard linearization of a nonlinear diffusion equation and avoids the explicit Jacobian required by Newton's method \cite{LangtangenLinge2017}. Picard iteration is generally more robust and simpler to implement, although its convergence is only linear and may become slow for very strong temperature dependence.

\subsubsection{Canonical five-point equation}

Substituting Eqs.~(\ref{e.radial_fd_flux}), (\ref{e.axial_fd_flux}) and (\ref{e.backward_euler}) into Eq.~(\ref{e.thermal_axisym}), while evaluating the nonlinear coefficients at Picard iterate $(m)$, gives the canonical five-point equation, or FDM stencil

\begin{equation} \label{e.thermal_five_point}
 { a_P T_P^{n+1,(m+1)}
 =a_E T_E^{n+1,(m+1)}
 +a_W T_W^{n+1,(m+1)}
 +a_N T_N^{n+1,(m+1)}
 +a_S T_S^{n+1,(m+1)}
 +b_P }.
\end{equation}

The east and west radial coefficients are

\begin{eqnarray}
&& a_E=\frac{r_e}{r_i}
\frac{k_{r,e}^{(m)}}{\Delta r_P\Delta r_E}, \\
&& a_W=\frac{r_w}{r_i}
\frac{k_{r,w}^{(m)}}{\Delta r_P\Delta r_W},
\label{e.thermal_coeff_rw}
\end{eqnarray}

and the north and south axial coefficients are

\begin{eqnarray}
&& a_N=\frac{k_{z,n}^{(m)}}{\Delta z_P\Delta z_N}, \\
&& a_S=\frac{k_{z,s}^{(m)}}{\Delta z_P\Delta z_S}.
\label{e.thermal_coeff_ns}
\end{eqnarray}

The central coefficient is

\begin{equation} \label{e.thermal_coeff_p}
 { a_P=
 \frac{c_{p,P}^{(m)}}{\Delta t}
 +a_E+a_W+a_N+a_S },
\end{equation}

while the right-hand-side contribution is

\begin{equation} \label{e.thermal_rhs}
 { b_P=
 \frac{c_{p,P}^{(m)}}{\Delta t}T_P^n
 +p_P }.
\end{equation}

Here, $p_P$ is the volumetric electromagnetic heat-generation density
supplied for cell $P$ at the current coupled time step. It remains fixed
during all Picard iterations, while the temperature-dependent heat capacity
and thermal conductivities are updated from the current Picard temperature
field. Equations~(\ref{e.thermal_five_point})--(\ref{e.thermal_rhs}) define a sparse linear system at each Picard iteration,

\begin{equation} \label{e.thermal_linear_system}
 { \ten{A}^{(m)}\vect{T}^{n+1,(m+1)}=\vect{b}^{(m)} },
\end{equation}

where $\ten{A}^{(m)}$ is the five-point thermal operator assembled using the material properties evaluated from $\vect{T}^{n+1,(m)}$, while $\vect{T}^{n+1,(m+1)}$ contains the unknown updated temperatures at time level $n+1$. The stencil associated with an interior cell is therefore

\begin{equation} \label{e.thermal_stencil}
 { \begin{array}{ccccc}
 && -a_N && \\
 & -a_W & a_P & -a_E & \\
 && -a_S &&
 \end{array} },
\end{equation}
which corresponds to equation \ref{e.thermal_five_point}. The Picard iteration is continued until the relative change in temperature is below a certain tolerance as
\begin{equation} \label{e.picard_convergence}
 { \frac{\left\|\vect{T}^{n+1,(m+1)}-\vect{T}^{n+1,(m)}\right\|_{\infty}}
 {\max\left(T_{\rm scale},\left\|\vect{T}^{n+1,(m+1)}\right\|_{\infty}\right)}
 <\epsilon_T },
\end{equation}
where $\epsilon_T$ is the prescribed nonlinear tolerance, $T_{\rm scale}$ prevents division by a very small reference temperature. The symbol $\|\cdot\|_\infty$ denotes the infinity norm, which for the discrete temperature field is the largest absolute value among all mesh cells,
\begin{equation}
 { \left\|\vect{x}\right\|_\infty
 =\max_P |x_P| }.
\end{equation}
Thus, the numerator in Eq.~(\ref{e.picard_convergence}) represents the maximum temperature change at any cell between two consecutive Picard iterations. Once convergence is reached, $\vect{T}^{n+1}=\vect{T}^{n+1,(m+1)}$ is accepted and the calculation advances to the next time step.

{

\subsection{Axisymmetric mechanical model}
\label{sec:mechanical_model}

The mechanical domain is an annular cross-section $\Omega=[r_{\rm inner},r_{\rm outer}]\times[z_{\rm bottom},z_{\rm top}]$ expressed in cylindrical coordinates $(r,\theta,z)$. The radial, circumferential and axial coordinates are denoted by $r$, $\theta$ and $z$, respectively. The axisymmetry of coil imposes $\partial(\cdot)/\partial\theta=0$, and the circumferential displacement is thus set to zero. The remaining unknown displacement components are the radial displacement $u_r(r,z)$ and the axial displacement $u_z(r,z)$.

\subsubsection{Strain and constitutive equations}

The total radial, hoop and axial normal strains ($\varepsilon_{ii}^{tot}$) are denoted by $\varepsilon_{rr}$, $\varepsilon_{\theta\theta}$ and $\varepsilon_{zz}$, while $\gamma_{rz}$ denotes the engineering shear strain. Under axisymmetry, the strain--displacement relations are

\begin{eqnarray}
&& \varepsilon_{rr}=\frac{\partial u_r}{\partial r}, \\
&& \varepsilon_{\theta\theta}=\frac{u_r}{r}, \\
&& \varepsilon_{zz}=\frac{\partial u_z}{\partial z}, \\
&& \gamma_{rz}=\frac{\partial u_r}{\partial z}+\frac{\partial u_z}{\partial r}.
\label{e.axisymmetric_strains}
\end{eqnarray}

The radial, hoop, axial and shear stresses are denoted by $\sigma_{rr}$, $\sigma_{\theta\theta}$, $\sigma_{zz}$ and $\sigma_{rz}$. Using the incremental thermal strains $\alpha_r\Delta T$, $\alpha_\theta\Delta T$ and $\alpha_z\Delta T$, the reduced orthotropic constitutive relation is

\begin{equation} \label{e.constitutive_matrix}
 { \left[
\begin{array}{c}
\sigma_{rr}\\
\sigma_{\theta\theta}\\
\sigma_{zz}\\
\sigma_{rz}
\end{array}\right]
=
\left[
\begin{array}{cccc}
C_{11} & C_{12} & C_{13} & 0\\
C_{12} & C_{22} & C_{23} & 0\\
C_{13} & C_{23} & C_{33} & 0\\
0 & 0 & 0 & G_{rz}
\end{array}\right]
\left[
\begin{array}{c}
\varepsilon_{rr}-\varepsilon_{rT}\\
\varepsilon_{\theta\theta}-\varepsilon_{\theta T}\\
\varepsilon_{zz}-\varepsilon_{zT}\\
\gamma_{rz}
\end{array}\right] },
\end{equation}
where, $\varepsilon_{iT} = \alpha_i\Delta T$ is the free thermal expansion strain ($\varepsilon_{ii}^{th}$) in the respective $r$, $\theta$ and $z$ directions, $\alpha$ is the thermal expansion coefficient, and $\Delta T$ denotes the temperature change relative to the chosen mechanical reference state. In general, $\alpha$ is temperature dependent, hoewever, we have considered it constant for this initial case study, as it does not vary much at cryogenic temperatures of 4 K-90 K for REBCO tape layers \cite{NISTmaterials}. Also, notably the strain contributions thus is the mechanical strains ($\varepsilon_{ii}^{mech}$), which is the required quantity for stress equilibrium.

Following the orthotropic formulation of Jones \cite{Jones1999}, the
material axes $1$, $2$ and $3$ are identified with the radial, hoop and
axial directions, respectively, i.e.,
$1\equiv r$, $2\equiv\theta$ and $3\equiv z$. 

The common denominator appearing in the orthotropic stiffness coefficients is

\begin{equation}
 { \Delta =
 \frac{
 1
 -\nu_{r\theta}\nu_{\theta r}
 -\nu_{\theta z}\nu_{z\theta}
 -\nu_{zr}\nu_{rz}
 -2\nu_{\theta r}\nu_{z\theta}\nu_{rz}
 }
 {E_rE_\theta E_z} }.
\label{e.orthotropic_delta}
\end{equation}

The normal stiffness coefficients are then given by \cite{Jones1999}

\begin{eqnarray}
&& C_{11}
 =\frac{1-\nu_{\theta z}\nu_{z\theta}}
 {E_\theta E_z\,\Delta},
\\
&& C_{22}
 =\frac{1-\nu_{rz}\nu_{zr}}
 {E_r E_z\,\Delta},
\\
&& C_{33}
 =\frac{1-\nu_{r\theta}\nu_{\theta r}}
 {E_r E_\theta\,\Delta}.
\label{e.orthotropic_normal_C}
\end{eqnarray}

The coupling coefficients are

\begin{eqnarray}
&& C_{12}
 =\frac{\nu_{\theta r}
 +\nu_{zr}\nu_{\theta z}}
 {E_\theta E_z\,\Delta}
 =\frac{\nu_{r\theta}
 +\nu_{z\theta}\nu_{rz}}
 {E_r E_z\,\Delta},
\\
&& C_{13}
 =\frac{\nu_{zr}
 +\nu_{\theta r}\nu_{z\theta}}
 {E_\theta E_z\,\Delta}
 =\frac{\nu_{rz}
 +\nu_{r\theta}\nu_{\theta z}}
 {E_r E_\theta\,\Delta},
\\
&& C_{23}
 =\frac{\nu_{z\theta}
 +\nu_{r\theta}\nu_{zr}}
 {E_r E_z\,\Delta}
 =\frac{\nu_{\theta z}
 +\nu_{\theta r}\nu_{rz}}
 {E_r E_\theta\,\Delta}.
\label{e.orthotropic_coupling_C}
\end{eqnarray}

The shear stiffness coefficients, where  $C_{44}$ and $C_{66}$ are ignored due to axisymmetry, are

\begin{equation}
 { C_{44}=G_{\theta z},
 \qquad
 C_{55}=G_{rz},
 \qquad
 C_{66}=G_{r\theta} }.
\label{e.orthotropic_shear_C}
\end{equation}

\subsubsection{Axisymmetric equilibrium and displacement equations}

The quasi-static radial and axial equilibrium equations are obtained from the divergence of the Cauchy stress tensor. Thus, the equilibrium equations are \cite{Jones1999}

\begin{eqnarray}
&& \frac{\partial\sigma_{rr}}{\partial r}
+\frac{\partial\sigma_{rz}}{\partial z}
+\frac{\sigma_{rr}-\sigma_{\theta\theta}}{r}+f_r=0, \\
&& \frac{\partial\sigma_{rz}}{\partial r}
+\frac{\partial\sigma_{zz}}{\partial z}
+\frac{\sigma_{rz}}{r}+f_z=0,
\label{e.axisymmetric_equilibrium}
\end{eqnarray}
Here, $f_r$ and $f_z$ are the radial and axial body force densities, respectively.
The thermal stress-coupling coefficients are denoted by $\beta_r$, $\beta_\theta$ and $\beta_z$. They are defined before their use in the displacement equations as

\begin{eqnarray}
&& \beta_r=C_{11}\alpha_r+C_{12}\alpha_\theta+C_{13}\alpha_z, \\
&& \beta_\theta=C_{12}\alpha_r+C_{22}\alpha_\theta+C_{23}\alpha_z, \\
&& \beta_z=C_{13}\alpha_r+C_{23}\alpha_\theta+C_{33}\alpha_z.
\label{e.beta_coefficients}
\end{eqnarray}

Substitution of equations~(\ref{e.axisymmetric_strains}) and (\ref{e.constitutive_matrix}) into equation~(\ref{e.axisymmetric_equilibrium}) gives the coupled radial displacement equation

\begin{equation} \label{e.radial_displacement_pde}
\begin{split}
 C_{11}\left(\frac{\partial^2u_r}{\partial r^2}
 +\frac{1}{r}\frac{\partial u_r}{\partial r}\right)
 -C_{22}\frac{u_r}{r^2}
 +G_{rz}\frac{\partial^2u_r}{\partial z^2} \qquad & \\
 +(C_{13}+G_{rz})\frac{\partial^2u_z}{\partial r\partial z}
 +\frac{C_{13}-C_{23}}{r}\frac{\partial u_z}{\partial z}
 &={} -f_r \\
 &\quad +\beta_r\frac{\partial T}{\partial r}
 +\frac{\beta_r-\beta_\theta}{r}\Delta T .
\end{split}
\end{equation}

The corresponding axial displacement equation is

\begin{equation} \label{e.axial_displacement_pde}
\begin{split}
 G_{rz}\left(\frac{\partial^2u_z}{\partial r^2}
 +\frac{1}{r}\frac{\partial u_z}{\partial r}\right)
 +C_{33}\frac{\partial^2u_z}{\partial z^2} \qquad & \\
 +(C_{13}+G_{rz})\frac{\partial^2u_r}{\partial r\partial z}
 +\frac{C_{23}+G_{rz}}{r}\frac{\partial u_r}{\partial z}
 &={} +f_z
 +\beta_z\frac{\partial T}{\partial z} .
\end{split}
\end{equation}

Equations~(\ref{e.radial_displacement_pde}) and (\ref{e.axial_displacement_pde}) show the complete one-way inputs received by the mechanical formulation at a given coupled iteration. Here, The electromagnetic contribution appears through $J_\theta B_z$ and $J_\theta B_r$, while the thermal contribution appears through the temperature gradients and the anisotropic free thermal strain.

\subsubsection{Mechanical boundary and material-interface conditions}
\label{sec:mechanical_bcs}

The mechanical domain contains both the homogenized pancake windings and the
G10 inter-pancake spacers. Each cell is assigned the corresponding local
elastic and thermal properties. The G10 spacers are therefore internal
material regions and not free mechanical boundaries. Explicit
pancake-to-pancake G10 spacers are also used in practical stacked REBCO
double-pancake magnets \cite{Hahn2010,Iwasa2015,Qu2017}.

The bottom surface is represented by a frictionless roller, whereas the top,
inner-radius and outer-radius surfaces are traction free. The external
conditions are

\begin{eqnarray}
&& z=z_{\mathrm{b}}:
\qquad u_z=0, \qquad \sigma_{rz}=0,
\nonumber\\
&& z=z_{\mathrm{t}}:
\qquad \sigma_{zz}=0, \qquad \sigma_{rz}=0,
\nonumber\\
&& r=r_{\mathrm{in}}:
\qquad \sigma_{rr}=0, \qquad \sigma_{rz}=0,
\nonumber\\
&& r=r_{\mathrm{out}}:
\qquad \sigma_{rr}=0, \qquad \sigma_{rz}=0.
\label{e.mechanical_bcs}
\end{eqnarray}

These are mixed displacement--traction conditions of the standard
thermoelastic boundary-value problem \cite{BoleyWeiner1960,Bower2010}.
In particular, the bottom surface is not clamped: only the axial displacement
is constrained, while radial expansion remains free. The shear-free condition
at this surface gives

\[
\left.\frac{\partial u_r}{\partial z}\right|_{z_{\mathrm{b}}}
=
-\left.\frac{\partial u_z}{\partial r}\right|_{z_{\mathrm{b}}}.
\]

One ghost-cell layer is used to impose the external conditions while retaining
the interior finite-difference structure \cite{LeVeque2007}. At the bottom
roller the axial displacement is imposed by antisymmetry,

\[
u_{z,i,0}=-u_{z,i,1},
\]

while the shear condition gives

\[
u_{r,i,0}
=
u_{r,i,1}
+
\Delta z_1
\left(\frac{\partial u_z}{\partial r}\right)_{i,1}.
\]

At the traction-free surfaces, the normal condition must be applied to the
full thermoelastic stress. Thus, the axial strain required at the top surface
and the radial strain required at either radial free surface are

\begin{eqnarray*}
\widehat{\varepsilon}_{zz}^{\,\mathrm{t}}
&=&
\alpha_z\Delta T
-\frac{
C_{13}(\varepsilon_{rr}-\alpha_r\Delta T)
+C_{23}(\varepsilon_{\theta\theta}-\alpha_\theta\Delta T)}
{C_{33}},
\\
\widehat{\varepsilon}_{rr}^{\,\mathrm{f}}
&=&
\alpha_r\Delta T
-\frac{
C_{12}(\varepsilon_{\theta\theta}-\alpha_\theta\Delta T)
+C_{13}(\varepsilon_{zz}-\alpha_z\Delta T)}
{C_{11}},
\end{eqnarray*}

where the superscripts $\mathrm{t}$ and $\mathrm{f}$ denote the top and a
radial free surface, respectively. The material coefficients in these
relations are always those of the material adjacent to the surface; hence,
a radial free surface crossing a G10 spacer uses the G10 properties.

For a displacement component $\phi$, the corresponding ghost value is
constructed from the prescribed normal derivative using

\[
\left.\frac{\partial\phi}{\partial n}\right|_N
\simeq
\frac{\phi_{N+1}-\phi_{N-1}}{d_N},
\qquad
\phi_{N+1}
=
\phi_{N-1}
+
d_N
\left.\frac{\partial\phi}{\partial n}\right|_N ,
\]

where $d_N$ is the actual distance between the two points straddling the
boundary. This form also accommodates the non-uniform axial spacing used for
the G10 layers.

The temperature change may vary spatially throughout the winding stack,
$\Delta T=\Delta T(r,z)$. Its gradients contribute to the thermoelastic
loading in the interior equations, while the local thermal strain also enters
the traction conditions at the external boundaries. A spatially uniform
temperature change represents the limiting case in which the interior
temperature-gradient terms vanish, although a finite free thermal strain
remains. This limiting case illustrates why the non-homogeneous thermal part
of the traction conditions must be included in the mechanical solution rather
than applied only after convergence.

To retain a linear matrix--vector operator, the displacement-dependent
homogeneous boundary relations are included in $\bm{A}$, whereas the known
thermal boundary contribution is transferred to the right-hand side,

\begin{equation}
\bm{A}\vect{x}
=
\vect{b}^{\,\mathrm{int}}
-
\vect{q}^{\,\mathrm{bc}},
\label{e.boundary_source}
\end{equation}

where $\vect{b}^{\,\mathrm{int}}$ contains the Lorentz-force and interior
thermoelastic terms, and $\vect{q}^{\,\mathrm{bc}}$ is the contribution
generated by the non-homogeneous thermal traction conditions. This
decomposition is valid for both uniform and spatially varying temperature
fields. This separation preserves the
linearity of the matrix-vector product required by the Krylov solver
\cite{LeVeque2007,Saad2003}.

\subsubsection{Pancake-G10 interfaces}

Perfect bonding is assumed between each winding and the adjacent G10 spacer.
Let $\Gamma_{\mathrm{c/G10}}$ denote such an interface, with superscripts
$\mathrm{c}$ and $\mathrm{G10}$ referring to the coil and G10 regions,
respectively. Displacement and traction continuity require

\begin{eqnarray}
&&
u_r^{\mathrm{c}}=u_r^{\mathrm{G10}},
\qquad
u_z^{\mathrm{c}}=u_z^{\mathrm{G10}},
\nonumber\\
&&
\sigma_{zz}^{\mathrm{c}}=\sigma_{zz}^{\mathrm{G10}},
\qquad
\sigma_{rz}^{\mathrm{c}}=\sigma_{rz}^{\mathrm{G10}} .
\label{e.g10_interface}
\end{eqnarray}

For a horizontal interface the normal is in the $z$ direction, and therefore
$\sigma_{zz}$ and $\sigma_{rz}$ are the traction components that must be
continuous \cite{Bower2010,MittelstedtBecker2007}. The radial and hoop
stresses are not required to be continuous across this interface.
Similarly, displacement continuity does not imply strain continuity:
different winding and G10 stiffnesses and thermal-expansion coefficients can
produce different strain gradients on the two sides of the interface.

\subsubsection{Cell-centred finite-difference discretization}
\label{sec:mechanical_fdm}

The mechanical equations are discretized on the same cell-centred grid and
using the same $P$, $E$, $W$, $N$ and $S$ notation introduced for the thermal
model. The displacement components $u_r$ and $u_z$ are stored at the cell
centres. In the present mechanical implementation the radial spacing
$\Delta r$ is uniform, whereas the axial cell height $\Delta z_P$ may vary
between cells. This allows the pancake winding and the G10 spacers to be
resolved with different axial cell sizes while retaining a structured
rectangular mesh. One ghost-cell layer is added outside the physical domain
to impose the external mechanical boundary conditions
\cite{LeVeque2007,HosseinverdiFasel2020}.

For a generic cell-centred variable $\phi$, the radial first- and
second-derivative operators at an interior cell are

\begin{equation}
\delta_r\phi_P
=
\frac{\phi_E-\phi_W}{2\Delta r},
\qquad
\delta_{rr}\phi_P
=
\frac{\phi_E-2\phi_P+\phi_W}{\Delta r^2}.
\label{e.mech_radial_operators}
\end{equation}

In the axial direction, the spacing can be non-uniform. Using the
centre-to-centre distances $\Delta z_N$ and $\Delta z_S$ introduced for the
thermal grid, the three-point second-order first derivative is

\begin{equation}
\begin{split}
\delta_z\phi_P
={}&
-\frac{\Delta z_N}
{\Delta z_S(\Delta z_S+\Delta z_N)}\phi_S
+\frac{\Delta z_N-\Delta z_S}
{\Delta z_S\Delta z_N}\phi_P
\\
&+
\frac{\Delta z_S}
{\Delta z_N(\Delta z_S+\Delta z_N)}\phi_N .
\end{split}
\label{e.mech_nonuniform_dz}
\end{equation}

For equal axial spacing,
$\Delta z_N=\Delta z_S=\Delta z$, equation~(\ref{e.mech_nonuniform_dz})
reduces to the usual centred expression
$(\phi_N-\phi_S)/(2\Delta z)$.

The corresponding non-uniform second derivative is

\begin{equation}
\delta_{zz}\phi_P
=
\frac{2}{\Delta z_N+\Delta z_S}
\left[
\frac{\phi_N-\phi_P}{\Delta z_N}
-
\frac{\phi_P-\phi_S}{\Delta z_S}
\right].
\label{e.mech_nonuniform_dzz}
\end{equation}

In the homogeneous parts of the winding, these operators reduce to the
standard centred finite differences. At a winding--G10 interface,
displacement remains continuous but its axial derivative can change because
of the different elastic and thermal properties. Therefore, axial
first-derivative and mixed-derivative stencils used for strain evaluation are
restricted to cells belonging to the same material. A second-order
one-sided three-point formula is used at the first or last cell of a material
region whenever three same-material points are available.

At the external radial boundaries, the forward and backward second-order
first derivatives remain

\begin{equation}
\delta_r^{+}\phi_{1,j}
=
\frac{-3\phi_{1,j}+4\phi_{2,j}-\phi_{3,j}}{2\Delta r},
\qquad
\delta_r^{-}\phi_{N_r,j}
=
\frac{3\phi_{N_r,j}-4\phi_{N_r-1,j}+\phi_{N_r-2,j}}
{2\Delta r}.
\label{e.mech_one_sided_r}
\end{equation}

For the non-uniform axial grid, the corresponding one-sided operators are
obtained from the quadratic interpolant through the first or last three cell
centres. At the lower boundary,

\begin{equation}
\begin{split}
\delta_z^{+}\phi_{i,1}
={}&
\frac{2z_1-z_2-z_3}
{(z_1-z_2)(z_1-z_3)}\phi_{i,1}
\\
&+
\frac{z_1-z_3}
{(z_2-z_1)(z_2-z_3)}\phi_{i,2}
+
\frac{z_1-z_2}
{(z_3-z_1)(z_3-z_2)}\phi_{i,3},
\end{split}
\label{e.mech_one_sided_z_bottom}
\end{equation}

while at the upper boundary,

\begin{equation}
\begin{split}
\delta_z^{-}\phi_{i,N_z}
={}&
\frac{z_{N_z}-z_{N_z-1}}
{(z_{N_z-2}-z_{N_z-1})
 (z_{N_z-2}-z_{N_z})}\phi_{i,N_z-2}
\\
&+
\frac{z_{N_z}-z_{N_z-2}}
{(z_{N_z-1}-z_{N_z-2})
 (z_{N_z-1}-z_{N_z})}\phi_{i,N_z-1}
\\
&+
\frac{2z_{N_z}-z_{N_z-2}-z_{N_z-1}}
{(z_{N_z}-z_{N_z-2})
 (z_{N_z}-z_{N_z-1})}\phi_{i,N_z}.
\end{split}
\label{e.mech_one_sided_z_top}
\end{equation}

Equations~(\ref{e.mech_one_sided_z_bottom}) and
(\ref{e.mech_one_sided_z_top}) reduce to the familiar
$(-3,4,-1)/(2\Delta z)$ and $(3,-4,1)/(2\Delta z)$ formulas for a uniform
axial mesh.

Pure radial second derivatives are retained in centred form at
boundary-adjacent cells, including when they use a physically constructed
ghost value. First and mixed derivatives are instead evaluated with the
one-sided expressions above whenever a centred stencil would use a ghost
value generated by the cross-component shear condition. This avoids a
circular dependence between perpendicular boundary closures while retaining
second-order spatial accuracy.

\subsubsection{Conservative axial traction discretization}

The axial terms require additional treatment because the elastic coefficients
change discontinuously between a pancake and a G10 spacer. Rather than
directly expanding terms such as
$G_{rz}\partial^2u_r/\partial z^2$ or
$C_{33}\partial^2u_z/\partial z^2$ across a material interface, the axial
stress divergence is evaluated from common face tractions. This conservative
form is applied throughout the internal axial domain so that the same
discretization is used in homogeneous regions and at material interfaces.

Consider the north face between cells $P$ and $N$. The distances from the two
cell centres to the common face are

\[
h_P=\frac{\Delta z_P}{2},
\qquad
h_N=\frac{\Delta z_N^{\,\rm cell}}{2},
\]

where $\Delta z_P$ and $\Delta z_N^{\,\rm cell}$ are the corresponding cell
heights. The effective mechanical face coefficients for shear and axial normal deformation are

\begin{equation}
K_{rz,n}
=
\left(
\frac{h_P}{G_{rz,P}}
+
\frac{h_N}{G_{rz,N}}
\right)^{-1},
\qquad
K_{zz,n}
=
\left(
\frac{h_P}{C_{33,P}}
+
\frac{h_N}{C_{33,N}}
\right)^{-1}.
\label{e.mech_face_coefficients}
\end{equation}

The same definitions are used at the south face. These
compliance-weighted coefficients account simultaneously for changes in
material stiffness and axial cell height.

The shear traction at the north face is

\begin{equation}
\sigma_{rz,n}
=
K_{rz,n}
\left[
u_{r,N}-u_{r,P}
+
(h_P+h_N)
\left(\delta_r u_z\right)_n
\right],
\label{e.mech_srz_face}
\end{equation}

where $(\delta_r u_z)_n$ is the radial derivative interpolated to the common
face.

For the elastic part of the axial normal traction, define

\[
\Psi_m
=
C_{13,m}\varepsilon_{rr,n}
+
C_{23,m}\varepsilon_{\theta\theta,n},
\qquad m=P,N,
\]

where the radial and hoop strains are evaluated at the common face. The
elastic normal traction is then

\begin{equation}
\sigma_{zz,n}^{\rm el}
=
K_{zz,n}
\left[
u_{z,N}-u_{z,P}
+
\frac{h_P\Psi_P}{C_{33,P}}
+
\frac{h_N\Psi_N}{C_{33,N}}
\right].
\label{e.mech_szz_face_el}
\end{equation}

The corresponding thermal contribution is

\begin{equation}
\sigma_{zz,n}^{\rm T}
=
-K_{zz,n}
\left[
\frac{h_P\beta_{z,P}\Delta T_P}{C_{33,P}}
+
\frac{h_N\beta_{z,N}\Delta T_N}{C_{33,N}}
\right],
\label{e.mech_szz_face_th}
\end{equation}

and the total normal traction is
$\sigma_{zz,n}=\sigma_{zz,n}^{\rm el}+\sigma_{zz,n}^{\rm T}$.
Equivalent expressions define $\sigma_{rz,s}$,
$\sigma_{zz,s}^{\rm el}$ and $\sigma_{zz,s}^{\rm T}$ at the south face.

The conservative axial derivatives are therefore

\begin{equation}
\left.
\frac{\partial\sigma_{rz}}{\partial z}
\right|_P
\simeq
\frac{\sigma_{rz,n}-\sigma_{rz,s}}{\Delta z_P},
\qquad
\left.
\frac{\partial\sigma_{zz}}{\partial z}
\right|_P
\simeq
\frac{\sigma_{zz,n}-\sigma_{zz,s}}{\Delta z_P}.
\label{e.mech_conservative_z}
\end{equation}

A single traction is thus shared by the cells on the two sides of every
internal face. At a winding--G10 interface this directly enforces the
traction-continuity conditions of equation~(\ref{e.g10_interface}); in a
homogeneous region with equal cell sizes it reduces to the conventional
centred finite-difference form.

\subsubsection{Final discrete mechanical equations}

Using the operators above, the radial equilibrium equation at cell $P$ is
written as

\begin{eqnarray}
&&
C_{11,P}
\left[
\frac{u_{r,E}-2u_{r,P}+u_{r,W}}{\Delta r^2}
+
\frac{u_{r,E}-u_{r,W}}{2r_P\Delta r}
\right]
-
C_{22,P}\frac{u_{r,P}}{r_P^2}
\nonumber\\
&&\quad
+
C_{13,P}\,
\delta_z\!\left(\delta_r u_z\right)_P
+
\frac{C_{13,P}-C_{23,P}}{r_P}
\,\delta_z u_{z,P}
+
\frac{\sigma_{rz,n}-\sigma_{rz,s}}{\Delta z_P}
\nonumber\\
&&\quad
=
-f_{r,P}
+
\beta_{r,P}\,\delta_r T_P
+
\frac{\beta_{r,P}-\beta_{\theta,P}}{r_P}\Delta T_P .
\label{e.radial_fd_full}
\end{eqnarray}

Here, $\delta_z(\delta_r u_z)_P$ is evaluated using the non-uniform
material-aware axial derivative described above. The conservative
$\sigma_{rz}$ term contains the remaining axial shear contribution, and
therefore replaces the explicit
$G_{rz}\partial^2u_r/\partial z^2$ part of the uniform-material stencil.

The final axial equation is

\begin{eqnarray}
&&
G_{rz,P}
\left[
\frac{u_{z,E}-2u_{z,P}+u_{z,W}}{\Delta r^2}
+
\frac{u_{z,E}-u_{z,W}}{2r_P\Delta r}
+
\delta_z\!\left(\delta_r u_r\right)_P
+
\frac{1}{r_P}\delta_z u_{r,P}
\right]
\nonumber\\
&&\quad
+
\frac{
\sigma_{zz,n}^{\rm el}
-
\sigma_{zz,s}^{\rm el}
}
{\Delta z_P}
=
f_{z,P}
+
b_{z,P}^{\rm T},
\label{e.axial_fd_full}
\end{eqnarray}

where the conservative axial thermal contribution is

\begin{equation}
b_{z,P}^{\rm T}
=
-
\frac{
\sigma_{zz,n}^{\rm T}
-
\sigma_{zz,s}^{\rm T}
}
{\Delta z_P}.
\label{e.axial_thermal_rhs}
\end{equation}

For a homogeneous material with constant $\beta_z$ and uniform axial spacing,
equation~(\ref{e.axial_thermal_rhs}) reduces to

\[
b_{z,P}^{\rm T}
=
\beta_z\,\delta_z T_P,
\]

and equations~(\ref{e.radial_fd_full}) and
(\ref{e.axial_fd_full}) reduce to the conventional centred finite-difference
forms of equations~(\ref{e.radial_displacement_pde}) and
(\ref{e.axial_displacement_pde}). Thus, the conservative formulation changes
neither the underlying mechanical equations nor the Lorentz-force loading;
it provides a consistent extension to the non-uniform axial grid and the
discontinuous winding--G10 material properties.

\subsubsection{Matrix-free BiCGSTAB solution}
\label{sec:mechanical_solver}

After spatial discretization, the radial and axial displacement unknowns are
collected into the global vector
$\vect{x}=[\vect{u}_r,\vect{u}_z]^{\mathrm{T}}$. The mechanical problem is
written in the form of Eq.~(\ref{e.boundary_source}), where $\bm{A}$ denotes
the discrete homogeneous elastic operator and $\vect{b}$ is the complete
right-hand side including the body-force, interior thermal and
non-homogeneous thermal boundary contributions.

The sparse matrix $\bm{A}$ is not assembled explicitly. Instead, its action
on a trial vector is evaluated directly from the finite-difference equations,
the homogeneous ghost-cell relations and the conservative axial face
tractions. The resulting algebraic operator is treated as non-symmetric
because of the cylindrical terms, coupling between the two displacement
components, non-uniform material interfaces and boundary closures.
The system is therefore solved using the Bi-Conjugate Gradient Stabilized
(BiCGSTAB) method \cite{vanDerVorst1992,Saad2003}.

A Jacobi preconditioner,
$\bm{M}=\mathrm{diag}(\bm{A})$, is used to improve the scaling of the two
displacement equations. For an internal cell $P$, its approximate radial and
axial diagonal entries, denoted by $d_{r,P}$ and $d_{z,P}$, are

\[
d_{r,P}
\simeq
\frac{2C_{11,P}}{\Delta r^2}
+
\frac{K_{rz,n}+K_{rz,s}}{\Delta z_P}
+
\frac{C_{22,P}}{r_P^2},
\]

and

\[
d_{z,P}
\simeq
\frac{2G_{rz,P}}{\Delta r^2}
+
\frac{K_{zz,n}+K_{zz,s}}{\Delta z_P}.
\]

Here, $K_{rz,n}$ and $K_{zz,n}$ are the north-face mechanical coefficients
defined in Eq.~(\ref{e.mech_face_coefficients}), while $K_{rz,s}$ and
$K_{zz,s}$ are their south-face counterparts. This face-based scaling
accounts for both the local material stiffness and the non-uniform axial cell
size, which is particularly useful at winding--G10 transitions.

At BiCGSTAB iteration $k$, the approximate solution is denoted by
$\vect{x}_k$ and the corresponding residual vector is

\[
\vect{r}_k
=
\vect{b}-\bm{A}\vect{x}_k ,
\]

where $k$ is the iteration index. The Euclidean vector norm is denoted by
$\|\cdot\|_2$. Starting from $\vect{x}_0=\vect{0}$, convergence is monitored
using the relative residual

\begin{equation}
R_k
=
\frac{\|\vect{r}_k\|_2}{\|\vect{b}\|_2}
=
\frac{\|\vect{b}-\bm{A}\vect{x}_k\|_2}
{\|\vect{b}\|_2}.
\label{e.relative_residual}
\end{equation}

The mechanical solve is considered converged when
$R_k<\varepsilon_{\mathrm{B}}$, where
$\varepsilon_{\mathrm{B}}$ is the prescribed BiCGSTAB convergence tolerance.
Standard breakdown checks are applied to the scalar denominators appearing
in the BiCGSTAB recurrences \cite{vanDerVorst1992,Saad2003}. After
convergence, the displacement field is used to evaluate the strains and
thermoelastic stresses from Eq.~(\ref{e.axisymmetric_strains}) and
Eq.~(\ref{e.constitutive_matrix}).

\subsection{Coupling of electromagnetic, thermal, and mechanical solvers}

We assume small strain, quasi-static equilibrium and linear orthotropic thermoelasticity in our model. The electromagnetic model supplies the azimuthal current density $J_\theta(r,z)$ and the radial and axial magnetic-flux-density components $B_r(r,z)$ and $B_z(r,z)$. The thermal model supplies the temperature $T(r,z)$, while $T_{\rm ref}$ denotes the stress-free reference temperature. The temperature change is written as $\Delta T=T-T_{\rm ref}$. The Lorentz body-force density is denoted by $\vect{f}=\vect{J}\times\vect{B}$, where $\vect{J}=J_\theta\vect{e}_\theta$ and $\vect{B}=B_r\vect{e}_r+B_z\vect{e}_z$. With this sign convention, the radial and axial mechanical inputs are

\begin{eqnarray}
&& f_r(r,z)=J_\theta(r,z)B_z(r,z), \\
&& f_z(r,z)=-J_\theta(r,z)B_r(r,z).
\label{e.lorentz_components}
\end{eqnarray}

Consequently, the mechanical model is coupled to the electromagnetic problem through $J_\theta$, $B_r$ and $B_z$, and to the thermal problem through $T$ and the temperature-dependent free thermal strain. The temperature change is considered from reference temperature $T_{ref}$ of 4.2 K at each time step. In the fully coupled problem, the electromagnetic constitutive law and critical current density depend on $\vect{B}$, $T$ and strain; these quantities should therefore be updated between nonlinear outer iterations, while the mechanical equations should be solved with coefficients frozen during each inner linear solve \cite{SiroisGrilli2015,PardoSoucFrolek2015,Barth2015}. However, at this initial stage of development, we consider quasi-static equilibrium in mechanical solver, which does not require temperature change between timesteps, and an overall update in temperature from $T_{ref}$ is enough for this case, as seen in the benchmark and results later. The coupling of thermal-electromagnetic model does consider update of temperature from previous time step, and it is shown in our previous work \cite{dadhichA2024SSTa}.

\subsection{Homogenized orthotropic properties}
\label{sec:mechanical_homogenization}

A resolved REBCO turn contains a metallic substrate, copper stabilizer, silver and buffer layers, the REBCO layer, and an inter-turn insulation layer. Explicitly resolving every constituent and interface is unnecessarily expensive when the layer period is much smaller than the dimensions over which the coil stress varies. The winding pack is therefore represented as an equivalent orthotropic continuum. This model reduction is consistent with the use of smeared orthotropic winding-pack properties in high-field REBCO magnet analyses and with long-wavelength homogenization of finely layered media \cite{Park2021,Backus1962}.

The homogenization of the electromagnetic and thermal properties, like electrical resistivity, thermal conductivity and thermal capacity is done in our previous work \cite{dadhichA2024SSTa}, and we use the same approach here as well. The mechanical behavior is considered orthotropic in our model and the homogenization process of these properties is discussed here.

Let $N$ denote the number of constituents in a representative turn or winding-pack unit cell. The thickness, total thickness and thickness fraction of constituent $m$ are denoted by $d_m$, $d=\sum_{m=1}^{N}d_m$ and $v_m=d_m/d$, respectively. The Young's modulus, shear modulus, Poisson ratio and coefficient of thermal expansion of constituent $m$ are denoted by $E_i^{(m)}$, $G_{ij}^{(m)}$, $\nu_{ij}^{(m)}$ and $\alpha_i^{(m)}$, where the directional indices $i$ and $j$ take values in $\{r,\theta,z\}$. For a property acting parallel to the constituent layers, the Voigt approximation (superscript $V$) corresponds to an iso-strain load path. For a property acting normal to the layers, the Reuss approximation (superscript $R$) corresponds to an iso-stress load path. These classical bounds provide transparent first estimates for a layered winding pack when detailed interface data are unavailable \cite{Hill1952,Backus1962}. The directional Voigt and Reuss estimates of the Young's and shear moduli are

\begin{eqnarray}
&& E_i^{\rm V}=\sum_{m=1}^{N}v_m E_i^{(m)}, \\
&& E_i^{\rm R}=\left(\sum_{m=1}^{N}\frac{v_m}{E_i^{(m)}}\right)^{-1}, \\
&& G_{ij}^{\rm V}=\sum_{m=1}^{N}v_m G_{ij}^{(m)}, \\
&& G_{ij}^{\rm R}=\left(\sum_{m=1}^{N}\frac{v_m}{G_{ij}^{(m)}}\right)^{-1}.
\label{e.voigt_reuss}
\end{eqnarray}

For a pancake winding in which the turn-to-turn stacking direction is radial, a first-order engineering assignment is

\begin{eqnarray}
&& E_r=E_r^{\rm R}, \qquad E_\theta=E_\theta^{\rm V}, \qquad E_z=E_z^{\rm V}, \\
&& G_{r\theta}=G_{r\theta}^{\rm R}, \qquad G_{rz}=G_{rz}^{\rm R}, \qquad G_{\theta z}=G_{\theta z}^{\rm V}.
\label{e.directional_homogenization}
\end{eqnarray}

Equation (\ref{e.directional_homogenization}) is an engineering approximation rather than a universal identity. It is appropriate when radial compression and the two radial shear modes traverse the layered turn-to-turn interfaces, whereas the hoop and axial normal load paths are predominantly parallel to the tape plane. Contact, partial bonding, voids, overbanding, co-winding and pancake-to-pancake interleaves may be included as additional constituents or represented by calibrated architecture factors. A more rigorous long-wavelength transversely isotropic baseline can be obtained from the Backus average \cite{Backus1962}; however, the Voigt-Reuss construction is retained here because it provides transparent upper and lower load-path estimates and can be applied independently in the three coil directions.

The three initial major Poisson ratios are denoted by $\nu_{r\theta}$, $\nu_{rz}$ and $\nu_{\theta z}$. They may be obtained from measured winding-pack data, unit-cell simulations, or, as a first estimate, a thickness-weighted average of the constituent values. The reciprocal Poisson ratios $\nu_{\theta r}$, $\nu_{zr}$ and $\nu_{z\theta}$ are not independent. They are defined through orthotropic reciprocity, which is required for a symmetric elastic compliance and stiffness tensor \cite{Ting1996}, as

\begin{eqnarray}
&& \frac{\nu_{r\theta}}{E_r}=\frac{\nu_{\theta r}}{E_\theta}, \qquad
\frac{\nu_{rz}}{E_r}=\frac{\nu_{zr}}{E_z}, \qquad
\frac{\nu_{\theta z}}{E_\theta}=\frac{\nu_{z\theta}}{E_z}.
\label{e.poisson_reciprocity}
\end{eqnarray}

The effective coefficients of thermal expansion in directions parallel to well-bonded layers are denoted by $\alpha_\theta$ and $\alpha_z$. Under the corresponding iso-strain assumption, they are stiffness weighted in accordance with energy-based thermoelastic composite estimates \cite{Schapery1968}. The radial coefficient $\alpha_r$ is taken as a thickness average under a series-like approximation. These definitions are

\begin{eqnarray}
&& \alpha_\theta(T)=
\frac{\sum_{m=1}^{N}v_m E_\theta^{(m)}(T)\alpha_\theta^{(m)}(T)}
{\sum_{m=1}^{N}v_m E_\theta^{(m)}(T)}, \\
&& \alpha_z(T)=
\frac{\sum_{m=1}^{N}v_m E_z^{(m)}(T)\alpha_z^{(m)}(T)}
{\sum_{m=1}^{N}v_m E_z^{(m)}(T)}, \\
&& \alpha_r(T)=\sum_{m=1}^{N}v_m\alpha_r^{(m)}(T).
\label{e.alpha_homogenized}
\end{eqnarray}

For a large cryogenic temperature excursion, a constant value of $\alpha_i$ is generally inadequate because the constituent contraction and polymer response can vary strongly with temperature. The free thermal strain in direction $i$ is therefore preferably evaluated from the integrated contraction

\begin{equation} \label{e.integrated_contraction}
 { \varepsilon_i^{\rm T}(T)=\int_{T_{\rm ref}}^{T}\alpha_i(\tau)\,\dd\tau },
\end{equation}

where $\tau$ is an integration temperature. For a sufficiently small temperature increment, equation~(\ref{e.integrated_contraction}) reduces to $\varepsilon_i^{\rm T}\simeq\alpha_i\Delta T$. The latter incremental form is used below to display the coupled displacement equations explicitly. The results below show that the model can describe thermal stresses in both small and large temperature change (ramp and quench respectively), and does not break down.

{
\subsection{Studied configuration}
}

\begin{table}
\centering
\caption{{Design, material and homogenized mechanical parameters of the
HTS metal-insulated nested stack} \cite{durochatM2024IES, NISTmaterials}.
{Copper thickness refers to the total thickness across the tape.}}
\label{param1}
\footnotesize
\renewcommand{\arraystretch}{1.10}

\begin{tabular}{@{}llll}

{\bf Nested stack parameters} & & & \\

Internal diameter
& 50 mm
& External diameter
& 216.2 mm \\

HTS1 pancakes
& 42
& HTS2 pancakes
& 44 \\

HTS1 turns per pancake
& 360
& HTS2 turns per pancake
& 260 \\

HTS1 external diameter
& 62.8 mm
& HTS2 internal diameter
& 201.2 mm \\

LTS magnetic field
& $\sim$15 T
& Total magnetic field at bore
& $\sim$40 T \\

Nominal current
& 231.2 A
& Initial ramp
& 0.1 A/s \\

Limiting voltage
& 2.5 V
& Contact resistance
& $10^{-6}$ $\Omega$m$^2$ \\[1mm]

{\bf Material dimensions} & & & \\

REBCO thickness
& 2 $\mu$m
& Substrate (Hastelloy) thickness
& 53 $\mu$m \\

Copper thickness
& 10 $\mu$m
& Turn-to-turn material thickness
& 30 $\mu$m \\

Tape width
& 6 mm
& Pancake spacer (G10)
& 0.5 mm \\[1mm]

{\bf Homogenized mechanical properties} & & & \\

Radial Young's modulus ($E_r$)
& 189.674 GPa
& Poisson's ratio ($\nu_{r\theta}=\nu_{rz}$)
& 0.2995 \\

Hoop/axial Young's modulus ($E_\theta=E_z$)
& 194.667 GPa
& Poisson's ratio ($\nu_{\theta r}=\nu_{zr}$)
& 0.3074 \\

Radial thermal expansion ($\alpha_r$)
& $11.643\times10^{-6}$ K$^{-1}$
& Poisson's ratio ($\nu_{\theta z}=\nu_{z\theta}$)
& 0.2995 \\

Hoop/axial thermal expansion ($\alpha_\theta=\alpha_z$)
& $11.310\times10^{-6}$ K$^{-1}$
& Radial-axial shear modulus ($G_{rz}$)
& 72.957 GPa \\

\end{tabular}
\end{table}
\normalsize

{We consider a user magnet design from the SuperEMFL project (basic design N1 for the 40 T magnet)} \cite{durochatM2024IES}. The LTS outsert is a 15 T/250 mm bore magnet from Oxford Instruments \cite{fazilleauP2024IES}. These summarized parameters are shown in Table \ref{param1}. The complete magnet system consists of LTS outsert, and {an HTS insert made of two nested stacks of metal-insulated pancake coils} with their own power (electrical) circuit, as shown in Figure \ref{crossSec}(a). {The useful bore will be slighly less than the inner diameter of 50 mm, due to the coil mandrel. Unless specified otherwise, we consider a contact resistance between turns of 10$^{-6}$ \Ohmmm, which could be achieved by metal insulation \cite{genotThesis2021}.} 

The HTS1 stack is connected to the HTS2 stack in series. The LTS stack provides a background field of around 15 T{, which is exactly 15 T at the bore center. We compute the exact local background magnetic field from the precise geometry and current density of the LTS windings that was provided by Oxford Instruments.} This whole system is initialized at 4.2 K, {and hence} liquid Helium temperature.

For the calculations, Theva APC tape is considered {(APC stands for Advanced Pinning Center)}, and its dimensions are given in Table \ref{param1}. The cross section of the tape is shown in Figure \ref{crossSec} (b). {Neighboring} turns are separated by a metal layer of Durnomag, which allows for radial currents between layers. {The electromagnetic and thermal properties for all the layers, like $\rho$, $C_v$, and $k$, are fully temperature dependent and their values are taken from {existing databases} \cite{NISTmaterials, durochatM2024IES}. {For $\rho$ of copper, we also take magneto-resistance into account.} The results in this paper consider the $J_c(B,T,\theta)$ of the Theva tape{, which is asymmetric \cite{senatoreC2024SST, pardoE2026RIE} (see figure \ref{JcBT} (a) and (b) )}}. The complete $J_c(B,T,\theta, \varepsilon)$ dependence allows to consider the decrease in the localized current density with magnetic field, temperature, and longitudinal strain (figure \ref{JcBT} (c)), that follows the irreversible relation as described in equations \ref{e.jc_strain_master_pw} and \ref{e.gmultiplier}. 

The constituent mechanical properties, including Young's modulus ($E$), Poisson's ratio ($\nu$), shear modulus ($G$), and coefficient of thermal expansion ($\alpha$), are obtained from available cryogenic material property data, including the NIST database \cite{NISTmaterials}. At this initial stage, the constituent elastic properties are assumed isotropic and temperature independent over the temperature range considered. The winding itself is not treated as isotropic: the individual material layers are homogenized in the $r$, $\theta$, and $z$ directions using the Voigt and Reuss approximations described above, resulting in an orthotropic constitutive model. For the present conductor composition, the resulting properties are nearly transversely isotropic in the $\theta$-$z$ plane, with $E_\theta=E_z$ and $\alpha_\theta=\alpha_z$, as summarized in Table~\ref{param1}. Using these homogenized values, the stiffness matrix coefficients are calculated as described in previous sections, which gives us final stiffness matrix for stress-strain relation, as shown is Equation \ref{e.constitutive_matrix}.

The constant mechanical property approximation is adopted here to isolate the initial thermo-mechanical coupling and is acceptable over a limited cryogenic temperature range (4.2 K- 90 K). Cryogenic elastic and thermal-expansion properties can vary with temperature \cite{NISTmaterials}; therefore, future work, particularly for large temperature transients during quench, should include temperature dependent elastic properties and thermal expansion. In particular, temperature-dependent thermal expansion or the corresponding integrated thermal contraction should be used when the mechanical reference state and the local temperature are separated by a large temperature interval.

{We consider that the magnet is first charged from 0 to a nominal current of 231.2 A at constant ramp of 0.1 A/s. Then, we wait 1156 s to enable relaxation of the current density, and hence the magnet is at stationary state. For the benchmark 
, we also simulate a fault damage, which can occur from local hotspots, manufacturing defects, or weak cooling. We assume that electrothermal quench initiates at a homogenized turn of the coil (group of 10 neighboring turns in one pancake) after achieving stationary state, and we set a strong reduction of $J_c$ down to 10 \% of the original, in order to avoid long prequench times. \R{The rest of the} assumptions and process for this study can be seen in our recent work \cite{dadhichA2026SUSTa}.

The current density and temperature profiles shown in section \ref{s.results} are for the cross section as shown in Figure \ref{crossSec} (a). \newline

\section{Benchmark of models}
\label{s.benckmark}

We have benchmarked the electromagnetic and thermal models multiple times in our previous works \cite{dadhichA2024SSTb, dadhichA2024SSTa, pardoE2026RIE}. The new thermal model is the classical implicit FDM, with the addition of non-linear solver of Picard iterations, and thus we do not show its validity in this paper to save space. \Rs{In addition, the considered nested stacks configuration is the same as our recent work, where the temperature rise is seen to be the same as this paper, and can be confirmed from Dadhich et al [dadhichA2026SUSTa].} The mechanical model is the newest addition to our software and we have benchmarked it in detail in this section. The FDM+BiCGSTAB based mechanical solver is compared with FEM model at CEA (combined with PEEC). We have also verified these results with Comsol FEM and Cast3DFM software internally, however, we mainly show the comparison with FEM from CEA to keep the discussion concise.

The benchmark is divided into 2 parts, where the stresses generated due to Lorentz forces and temperature rise are considered independently. This is done to check the accuracy of individual contributions of these multiphysical quantities seperately. The benchmark is done on a single full scale \R{metal-insulated pancake} coil, with same configuration of the coils from the nested stack (i.e. 360 turns, with same inner and outer radius as HTS1). The electromagnetic results were calculated by MEMEP, such as current density, and radial and axial forces, as seen in Figure \ref{benchmark_param} (a)-(c). These results acted as an input for both FDM and FEM solvers for comparison of stresses due Lorentz force. For thermal stresses benchmark, We assumed a damage in the coil at turns 41-50, which reduces the screening current density and increases the temperature in those parts (Figure \ref{benchmark_param} (d)). This case is simulated mainly with the purpose to generate complex non-uniform temperature profiles for the rigorous testing of the mechanical solver to compare stresses due to temperature change. The reduction of $J_c$ by mechanical strain was not considered in the benchmark, and only the mechanical properties were compared.

As seen in later results, considering General configuration for $J_c(\varepsilon)$ does not reduce $J_c$ significantly for the ramping of magnet, thus this assumption is considered \R{satisfactory} for the benchmark. \R{All benchmarked methods use the} same mesh, mechanical properties, and geometry. \R{In particular, they use 20 elements across the tape width and one element per turn across the thickness.}

\begin{figure}[tbp]
	\centering

{\includegraphics[trim=0 0 0 0,clip,width=16 cm]{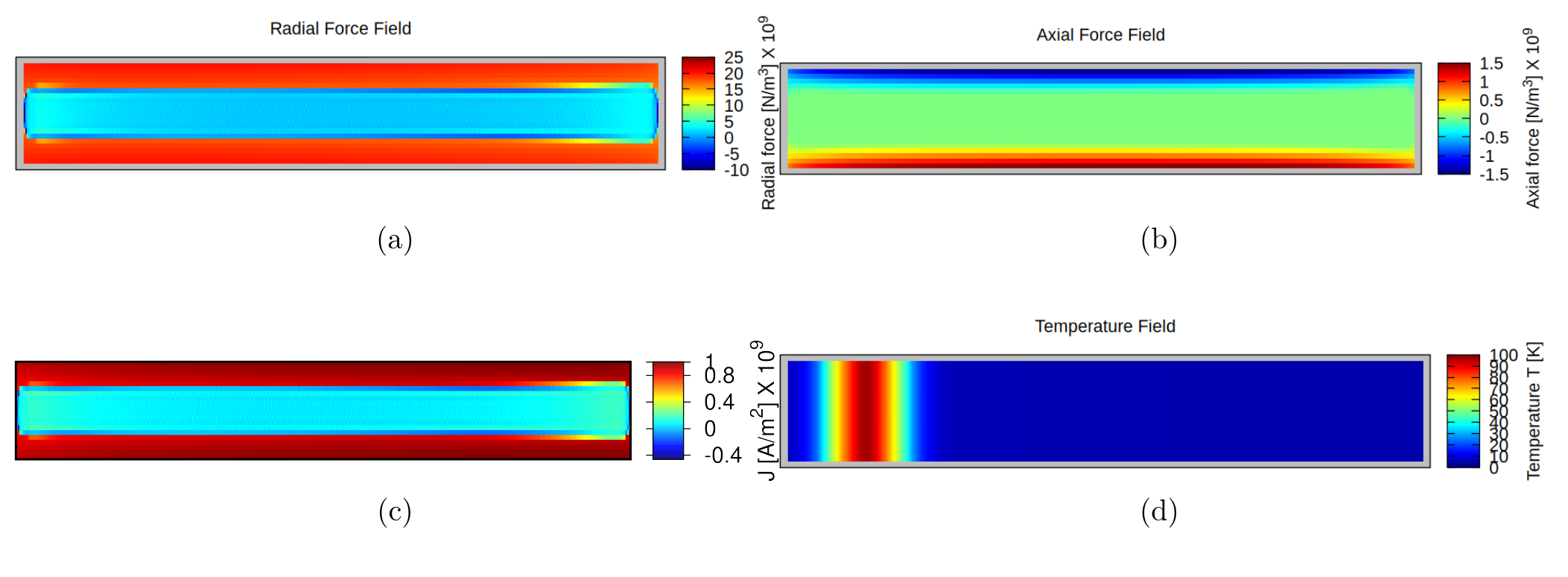}}%

\caption{Benchmark parameters, generated by MEMEP, shows non-uniform (a) radial forces, (b) axial forces, (c) screening currents, with no temperature change, to benchmark Lorentz stresses individually. (d) shows a special case of the coil where there is damage in the coil and temperature rises locally from the hotspot (used for benchmarking thermal stresses).  }
\label{benchmark_param}
\end{figure}

\subsection{Benchmark of Mechanical FDM with FEM: only Lorentz forces}

As seen in Figure \ref{benchmark_results_Lorentz}, we get good agreement between mechanical FDM and FEM results. The 1st and 2nd columns here present mechanical FDM results from IEE and CEA, respectively. In particular, we have very strong agreement in all different stresses, i.e., radial, hoop, axial and shear stresses between two methods. Thus we confirm the validity of the stresses generated by Lorentz forces in mechanical FDM.

\begin{figure*} [tbp]

{\includegraphics[trim=0 0 0 0,clip,width=16 cm]{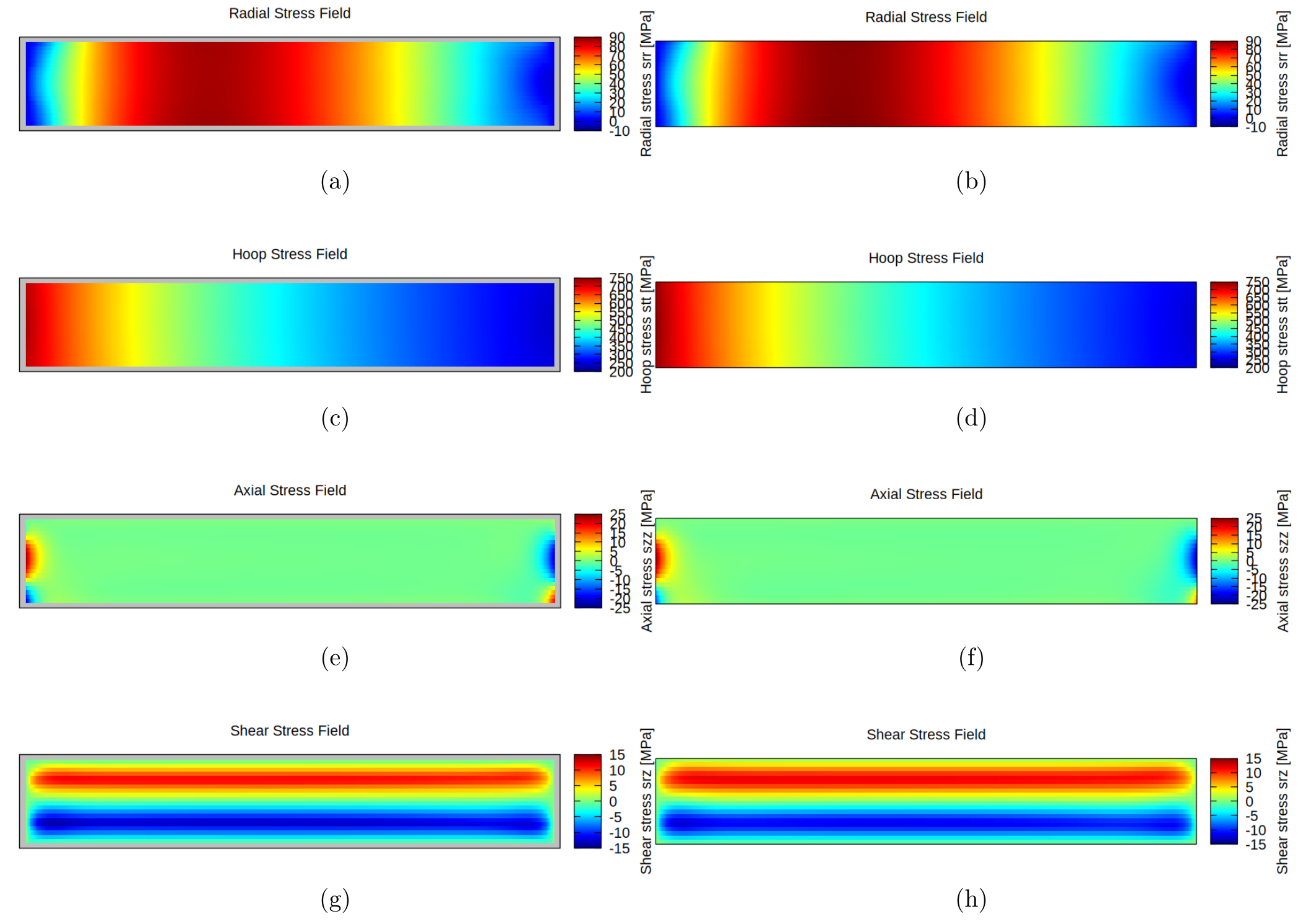}}%
 
\caption{Benchmark of Lorentz stresses with constant temperature between FDM and FEM. Results show good agreement between radial forces (a,b), hoop stresses (c,d), axial stresses(e,f), and shear stresses (g,h) for FDM (a,c,e,g) and FEM in PEECS (b,d,f,h). }
\label{benchmark_results_Lorentz}
\end{figure*}

\subsection{Benchmark of Mechanical FDM with FEM: only Temperature change} 

\R{We also} get good agreement between mechanical FDM and FEM results for thermal stresses, as seen in Figure \ref{benchmark_results_Thermal}. The thermal stresses show much different qualitative behavior, when compared with Lorentz stresses. The \R{largest} thermal stresses are \Rs{generally }compressive in nature; as opposed to Lorentz stresses, which are tensile. This is counter-intuitive, because thermal expansion is usually tensile. However, the general compressive behavior here is due to the surrounding tape windings, which constrain the local turns that try to expand, creating an overall compressive behavior.

\begin{figure*} [tbp]

{\includegraphics[trim=0 0 0 0,clip,width=17 cm]{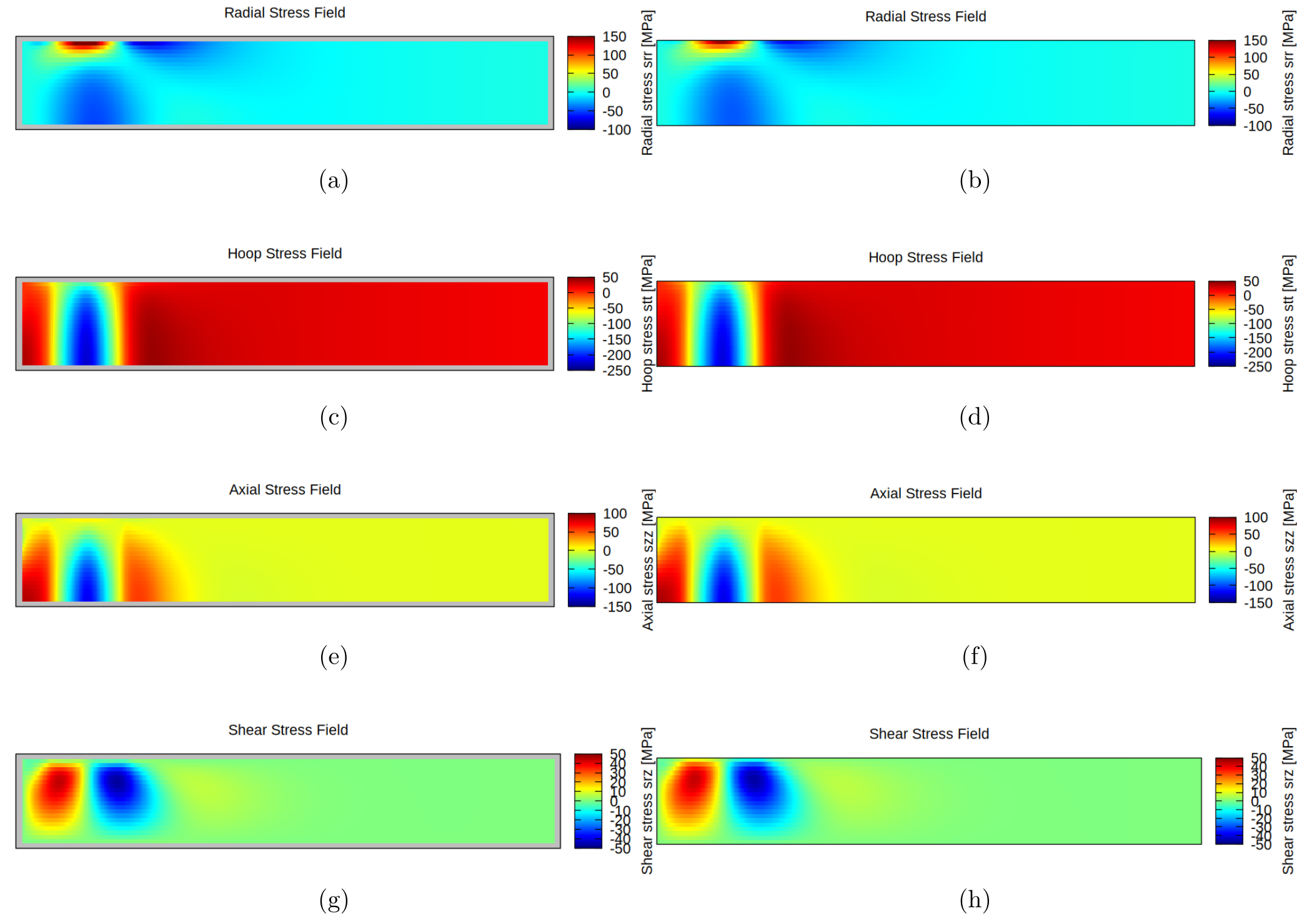}}%
 
\caption{Benchmark of Thermal stresses, ignoring Lorentz stresses, between FDM and FEM for a case with local hotspot. Results show good agreement between radial forces (a,b), hoop stresses (c,d), axial stresses(e,f), and shear stresses (g,h) for FDM (a,c,e,g) and FEM in PEECS (b,d,f,h). }
\label{benchmark_results_Thermal}
\end{figure*}

\section{Results and Discussion}
\label{s.results}

We have applied the coupled models to normal operation mode for the complete HTS nested stacks, as presented in the configuration section and Figure \ref{crossSec}, with field from LTS in background, giving us total of 40 T magnetic field at the bore. We have considered 2 cases of $I_c(\varepsilon)$ dependence, i.e., Conservative and General (Figure \ref{JcBT} (c)), as they closely resemble the experimental results from SuperEMFL deliverable \cite{SuperEMFLTARSIS}\Rs{, and Optimistic case is ignored for now until substantial experimental backing with strong REBCO is available for it}. 

\subsection{Case Study: Lorentz force effects at ramp up}

For the ramping up of the magnet in normal operating conditions (no artifical damage), we consider only the stresses due to Lorentz forces, and thermal gradients are ignored for this case study. We do so to as to individually assess their contributions for stress generation and $J_c$ reduction\Rs{, and thermal stresses are added in the next Case Study}. Firstly, it is seen from Figure \ref{rampUp}, that the magnet using Conservative case cannot ramp up fully without \R{experiencing thermal quench} \Rs{recurring significant losses}, as compared \R{to} the case with General $I_c$. Figure \ref{rampUp} (c) shows sharp rise in power loss at around 2300 s (very close to top of ramp) for Conservative case, which rises the temperature as well in the system (\R{Figure} \ref{rampUp} (\R{d})). The main reason for this is seen in Figure \ref{rampUp} (e) that near the inner radial surface of HTS1, \R{the decrease in $J_c$ due to strain is significant} (by almost 50 percent), as compared to \R{the} General case, where \R{the} drop is only around 10 percent. This drop in \R{$J_c$} is due to mechanical strain at this edge of more than 0.35 \% (Figure \ref{rampUp3} (g)) \R{that} is higher than the irreversible strain of this tape of 0.29 \%, where $J_c$ drops sharply for Conservative case (Figure \ref{JcBT} (c)). This power and temperature rise then reduces \R{$J_c$ through the} $J_c(B,T,\theta)$ \R{dependence} (Figure \ref{JcBT} (a)) and quench occurs. If not controlled, this quench can keep generating higher power losses and temperature runaway, \Rs{subsequently,} which can be highly damaging for the magnet. Thus, we apply the protective voltage limitation \R{of 2.5~V}, which decreases the current in the coil \R{during quench}, and hence \R{limits the total heat generation and maximum temperature.} \Rs{the magnet can keep operating at lower magnetic fields or switch of completely. }This is the reason we see the drops in current, magnetic field, and power loss in (Figure \ref{rampUp} (a-c)). As we assume adiabatic conditions for temperature, it keeps rising in Figure \ref{rampUp} (d). \R{Since the maximum temperature ($\sim$300~K) is well below the typical thermal damage temperature of 450~K, we do not expect direct thermal damage. Thus, permanent decrease of $J_c$ is only due to mechanical damage. Since it occurred at around 230~A, we expect that the magnet could still be charged at just below that current.}

\begin{figure*} [tbp]

     \centering
 {\includegraphics[trim=0 0 0 0,clip,width=17 cm]{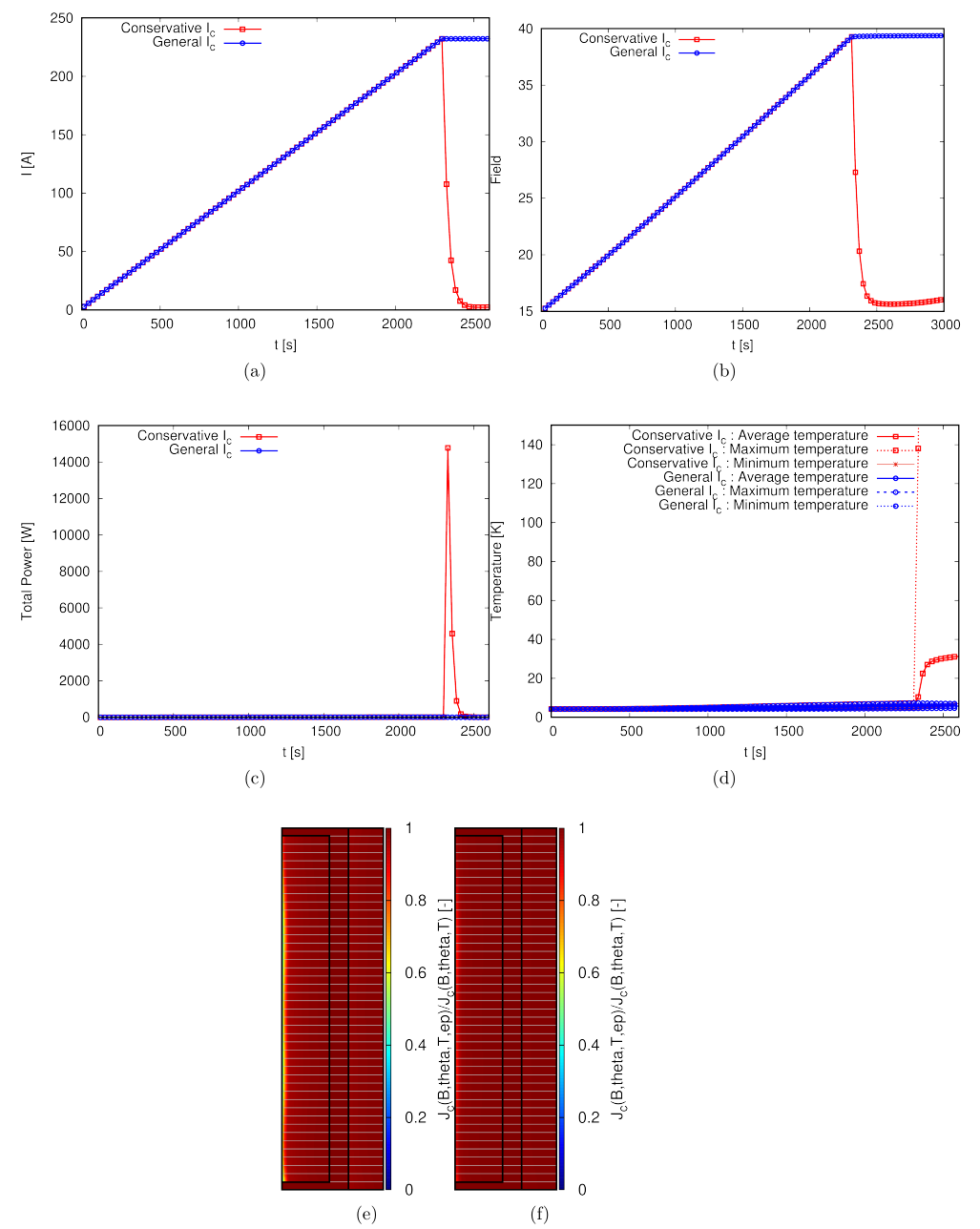}}%
    
  \caption{Ramping up behavior of magnet, under Conservative and General $I_c$ configurations, ignoring thermal stresses. The figures show (a) input current, (b) magnetic field at bore, (c) total power loss, (d) temperatures, and (e,f) normalized change in $J_c$ : $J_c(B,T,\theta, \varepsilon)$/$J_c(B,T,\theta)$, which shows decrease in $J_c$ at inner edge for Conservative $I_c$ case.}
  \label{rampUp}
\end{figure*}

\begin{figure*}[tbp]

	\centering
{\includegraphics[trim=0 0 0 0,clip,width=17 cm]{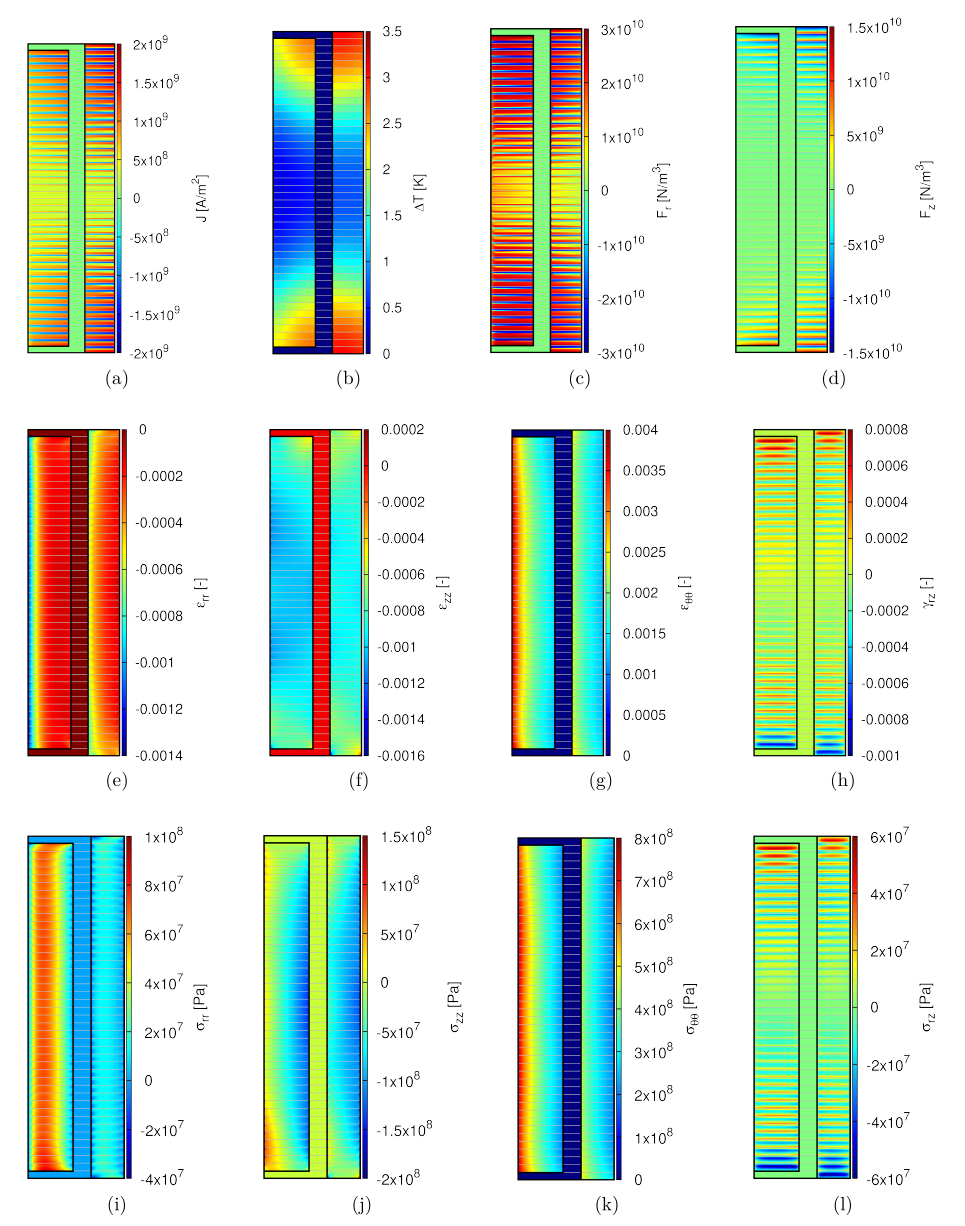}}%

  \caption{Ramping up behavior of magnet, under Conservative and General $I_c$ configurations, ignoring thermal stresses. The figures show (a) current density, (b) change in temperature, (c) radial Lorentz forces, (d) axial Lorentz forces, (e) radial strain, (f) axial strain, (g) hoop strain, (h) shear strain, (i) radial stress, (j) axial stress, (k) hoop stress, and (l) shear stress.  }
  \label{rampUp3}
\end{figure*}

Figure \ref{rampUp3} show\R{s} other multiphysical quantities for the ramping up of the magnet \R{at $t$=2300~s} \Rc{** Anang, check the particular time **}, which are \R{the} same for both Conservative and General cases (or with minimal differences\Rs{up to this time in simulation}). Figure \ref{rampUp3} (a)-(d) show electromagnetic and thermal properties, like screening current density (a), temperature change (b), radial forces (c), and axial forces (d), which are non-uniform in nature and synonymous with each other. These {\R non-uniformities are mainly} due the screening currents \Rs{and asymmetric $J_c$ behavior of Theva tape, as seen in Figure \ref{JcBT} (b), as well as, in our previous work [dadhichA2026SUSTa]}\Rc{** Now, you use symmetric Jc of Theva tape **}. These quantities also act as input for thermal and mechanical FDM, and it can be seen that the forces are quite high. These generate high mechanical stresses and strains in the system which cannot be ignored (Figure \ref{rampUp3} (e)-(l)). The maximum stresses are in \R{the} range of 100 MPa to 800 MPa, which can be damaging to \R{$J_c$ through $g_{\rm total}(\varepsilon,\varepsilon_{\rm max})$}\Rs{ the system electrically,}. However, they are still lower than the yield strength of REBCO (around 1000 MPa). It is seen that the stresses are higher at the innermost radius of nested stacks. The highest of these is hoop stress (upto 800 MPa), which also contributes to hoop strain generation and $J_c$ reduction, as discussed above. The axial stress and strain \R{are the} highest at the bottom and outer edges of the nested stacks (Figure \ref{rampUp3} (j) and (f)), which is due to considered bottom roller condition and \R{maximum axial magnetic field that generate high radial}\Rs{parallel field lines that generate high axial} Lorentz forces at vertical edges of the magnet. The shear stress is also \R{the} highest at the upper and lower edges, \R{due to screening currents}\Rs{due to the bottom roller condition, axial and radial stresses, and the perfectly bonded magnet consideration, which distributes the torsion and stresses via G10 to the full magnet}.

After $t=$2300~s \R{there is quench for the Conservative case}\Rs{ ramping up fails for the Conservative case, there is quench in the magnet}, and this behavior \Rs{in terms of different multiphysical quantities }is shown in Figures \ref{quench_con1} -\ref{quench_con3}. Figure \ref{quench_con1} \R{tells us} that the quench occurs on the innermost radius of the nested stack, which is due to the loss in local $J_c$ at that edge from high strain, as discussed above. \R{The asymmetry of the hot spot location is due to the roller boundary condition, which causes a slight asymmetry in $\varepsilon_{\theta\theta}$, and hence in $J_c$.}\Rs{A slight asymmetry for quench hotspot is from the combination of asymmetric $J_c$ of Theva, bottom roller mechanical boundary condition, and related generation of non-uniform stresses and power losses in different regions.} \R{Most of the AC loss is generated at the}\Rs{Mainly the power losses generate strongly from} innermost edge (Figure \ref{quench_con1} (d)), and they keep increasing with time, until the \R{current in the} magnet starts \R{decreasing}\Rs{ powering down} due to voltage limitation. This is due to reduction in $J_c$\Rs{ and $J$}, whose time evolution during quench can be seen in Figure \ref{quench_con1}\Rs{ (a) and} (b). This \R{highly}\Rs{uncontrollably} rises the temperature during the quench, up to almost 300~K, \R{although this should not directly create permanent damage in the magnet. However, high thermal gradients might create damage through thermal stress, although this has not been taken into account in this case study}.\Rs{which can be detrimental to magnet health, if this heat is not extracted.} 

Figures \ref{quench_con2} and \ref{quench_con3} show the time evolution of all strains \R{and stresses} during quench, which reduce with time as the current \R{decreases}\Rs{goes down} from voltage limitation. This shows the benefit of voltage limitation\R{. It reduces} the power losses and input current during quench for protection, \R{while it} also reduces Lorentz forces \R{(}and consequently stresses\R{)} related to these forces. It is important to note, however, that while the hoop strains are reduced during voltage limitation (as well as during ramping up and down of the magnet), the maximum hoop strain achieved ($\varepsilon_{max}$) is recorded in material's 'memory', which contributes to the irreversible $J_c$ behavior via PW model or specifically $g_W$ multiplier to $J_c(B,T,\theta)$ dependence. Hence, it is an important quantity to note for future simulations, as well as, experiments \R{or} magnet operations.

\begin{figure*} [tbp]

    \centering
	{\includegraphics[trim=0 0 0 0,clip,width=13 cm]{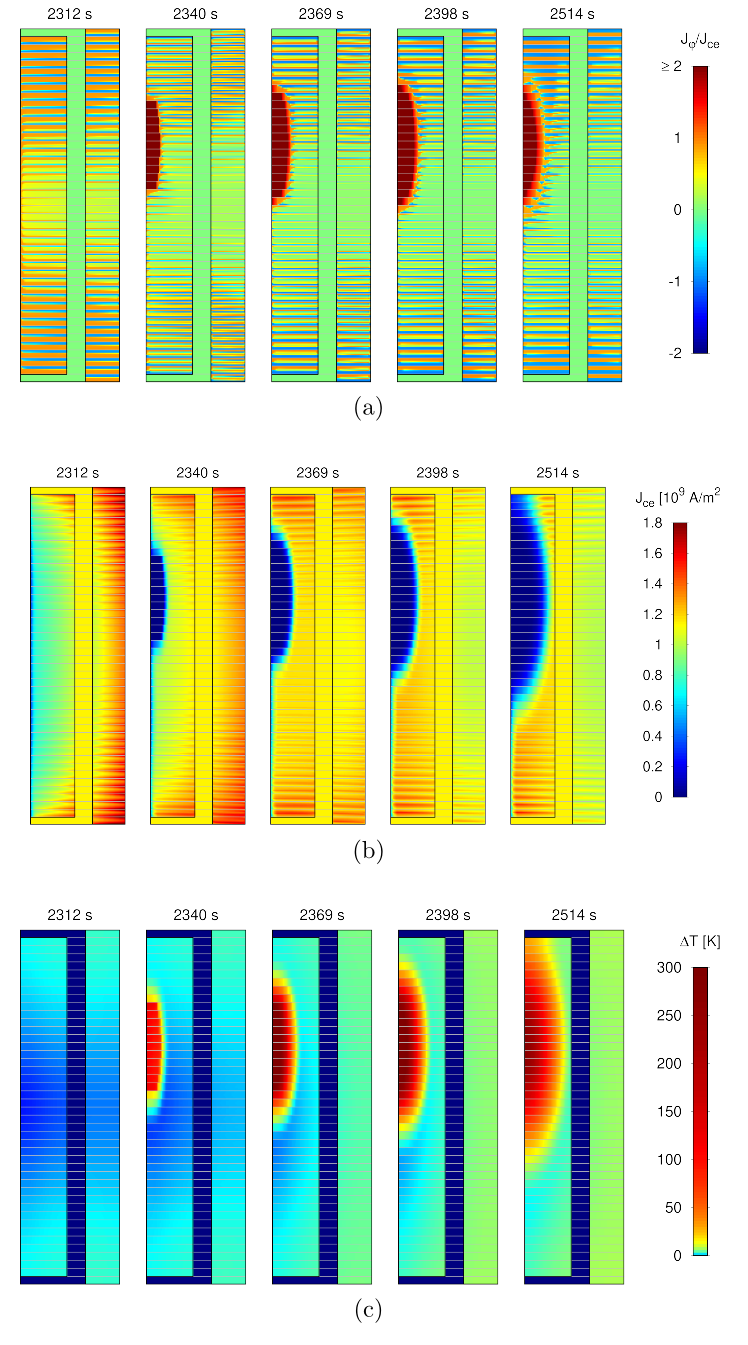}}%
    
  \caption{Time evolution of quench behavior of magnet, under Conservative $I_c$ configuration, ignoring thermal stresses. The figures show change in (a) normalized current density, (b) critical current density and (c) temperature. }
  \label{quench_con1}
\end{figure*}

\begin{figure*} [tbp]

    \centering
	{\includegraphics[trim=0 0 0 0,clip,width=10 cm]{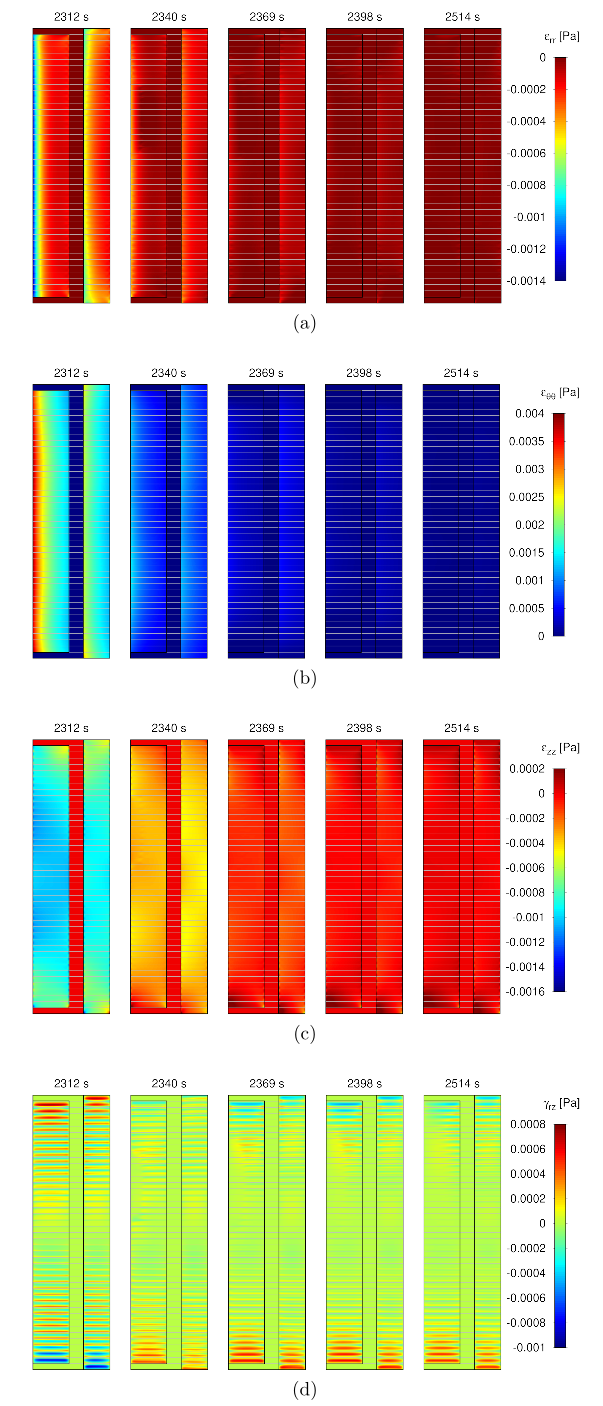}}%
    
  \caption{Time evolution of quench behavior of magnet, under Conservative $I_c$ configuration, ignoring thermal stresses. The figures show change in (a) Radial strain (b) Hoop strain (c) Axial strain, and (d) shear strain}
  \label{quench_con2}
\end{figure*}

\begin{figure*} [tbp]

    \centering
{\includegraphics[trim=0 0 0 0,clip,width=10 cm]{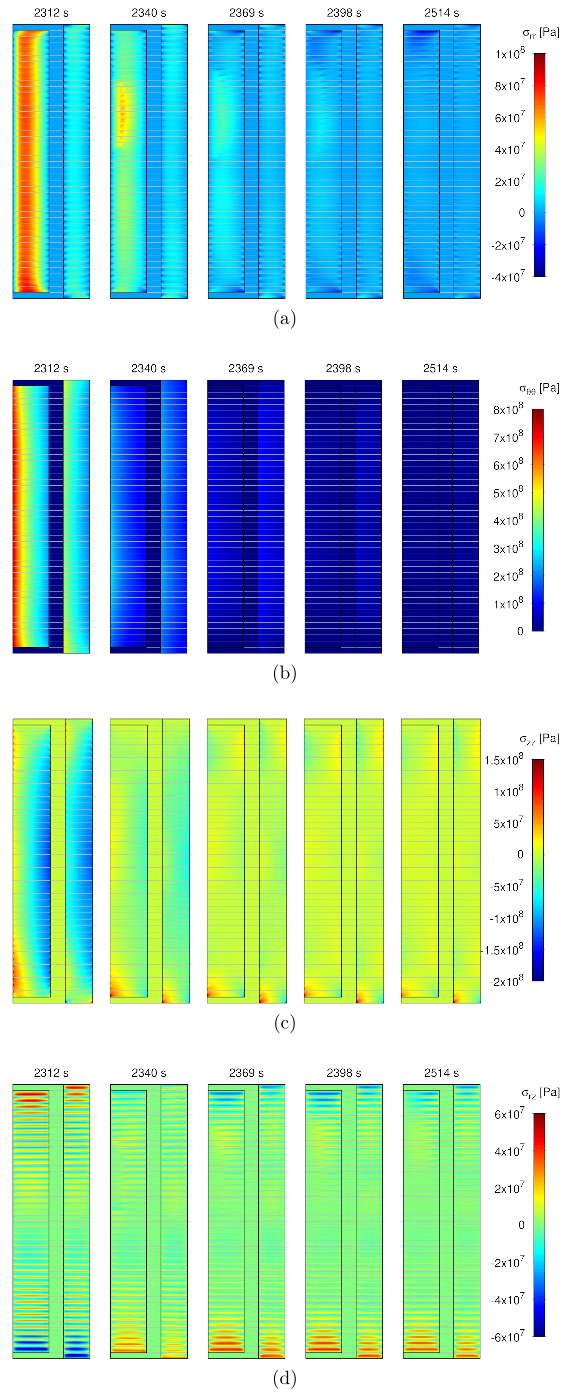}}%
    
  \caption{Time evolution of quench behavior of magnet, under Conservative $I_c$ configuration, ignoring thermal stresses. The figures show change in (a) Radial stress (b) Hoop stress (c) Axial stress, and (d) shear stress.}
  \label{quench_con3}
\end{figure*}

\section{Conclusion}

A coupled electromagnetic, thermal and mechanical framework, based on variational principles and FDM, has been developed for the metal-insulated nested REBCO insert considered for the SuperEMFL 40 T class magnet. The benchmark of the FDM based models show good agreement for Lorentz and thermal stresses generated by FEM. The results demonstrate that the Lorentz forces generate very high stresses, and the magnet quenches even during ramping up, when considered Conservative $J_{\rm c}(B,T,\vartheta,\varepsilon)$ dependence (consistent with SuperEMFL measurements). Localized heating can produce significant mechanical stresses even in the absence of Lorentz loading, since constrained thermal expansion generates compressive and tensile strain regions around the thermal disturbance. Thus, the samples should be chosen carefully for the windings and should portray atleast General $J_{\rm c}(B,T,\vartheta,\varepsilon)$ dependence for the normal operation of the magnet.

The developed model therefore provides a unified tool for studying the interaction between electromagnetic loading, thermal gradients, mechanical limits and quench in large nested REBCO inserts. \R{Proposed future work could be on}\Rs{work will include} improved temperature-dependent mechanical properties, more detailed conductor and interface damage descriptions, and further experimental validation of the strain-dependent $J_{\rm c}$ response.

\section{Acknowledgements}

We acknowledge Oxford Instruments for providing details on the cross-section of the LTS outsert. This {work} has received funding from the European Union's Horizon 2020 research and innovation programme under grant agreement No 951714 (superEMFL), and the Slovak Republic from projects APVV-24-0654 and VEGA 2/0098/24. Research partially funded by the EU NextGenerationEU through the Recovery and Resilience Plan for Slovakia under the project No. 09I04-03-V02-00039. Part of the research results were obtained using the computational resources procured in the national project National competence centre for high-performance computing (project code: 311070AKF2) funded by the European Regional Development Fund, EU Structural Funds Informatization of society, Operational Program Integrated Infrastructure. Any dissemination of results reflects only the authors' view and the European Commission is not responsible for any use that may be made of the information it contains. AD acknowledges the Schwarz Fund from the Slovak Academy of Sciences.

\end{document}